\documentclass{aa}  
\usepackage{graphicx}
\usepackage{txfonts}
\usepackage{xspace}
\usepackage{natbib}

\newcommand{\RV}{$R_V$\xspace} 
\newcommand{\Rmono}{$R_{5495}$\xspace} 
\newcommand{\Amono}{$A_{5495}$\xspace} 
 
\newcommand{\bump}{2175~\AA~feature\xspace}

\usepackage{listings}
\usepackage{xcolor}
\usepackage{float}

\usepackage{newfloat}
\DeclareFloatingEnvironment[fileext=lop,placement={tb},name=Listing]{lstfloat}

\floatname{lstfloat}{Listing}

\definecolor{codegreen}{rgb}{0,0.6,0}
\definecolor{codegray}{rgb}{0.5,0.5,0.5}
\definecolor{codepurple}{rgb}{0.58,0,0.82}
\definecolor{backcolour}{rgb}{0.95,0.95,0.92}

\lstdefinestyle{mystyle}{
    backgroundcolor=\color{backcolour},   
    commentstyle=\color{codegreen},
    keywordstyle=\color{magenta},
    numberstyle=\tiny\color{codegray},
    stringstyle=\color{codepurple},
    basicstyle=\ttfamily\tiny,
    breakatwhitespace=false,         
    breaklines=true,                 
    captionpos=b,                    
    keepspaces=true,                 
    numbersep=5pt,                  
    showspaces=false,                
    showstringspaces=false,
    showtabs=false,                  
    tabsize=2
}

\defcitealias{1999PASP..111...63F}{F99}
\newcommand{\fitz}{\citetalias{1999PASP..111...63F}\xspace}
\defcitealias{2014AA...564A..63M}{MA14}
\newcommand{\maiza}{\citetalias{2014AA...564A..63M}\xspace}

\newcommand{\Msun}{M$_{\odot}$\xspace}

\usepackage{hyperref}
\hypersetup{
    colorlinks=true,
    citecolor=blue,
}
\usepackage[capitalise]{cleveref}

\begin{document}

\title{Extinction towards the cluster R136 in the Large Magellanic Cloud}
\subtitle{An extinction law from the near-infrared to the ultraviolet}
 
   \author{Sarah A. Brands \inst{1}
          \and 
          Alex de Koter  \inst{1,3}
          \and 
          Joachim M. Bestenlehner \inst{2}
          \and 
          Paul A. Crowther \inst{2}
          \and 
          Lex Kaper       \inst{1}
          \and 
          Saida M. Caballero-Nieves   \inst{4}
          \and 
          G\"{o}tz Gr\"{a}fener \inst{5}
          }

\institute{ 
          Astronomical Institute Anton Pannekoek, 
          Amsterdam University,  
          Science Park 904, 1098~XH, 
          Amsterdam, The Netherlands\newline
          \email{s.a.brands@uva.nl}
          \and 
          Department of Physics and Astronomy
          University of Sheffield,
          Sheffield, S3~7RH, 
          United Kingdom
          \and 
          Institute of Astrophysics,
          KU Leuven, 
          Celestijnenlaan 200D, 
          3001, Leuven, Belgium
          \and 
          Embry-Riddle Aeronautical University,
          Department of Physical Science,
          1 Aerospace Blvd,
          Daytona Beach, FL 32114
         \and 
         Argelander-Institut f\"ur Astronomie, 
         Universit\"at Bonn, Auf dem H\"ugel 71, 
         53121 Bonn, Germany
         }

\date{Received 23 November, 2022 / Accepted 3 March 2023}

  \abstract
 {The cluster R136 in the giant star-forming region 30 Doradus in the Large Magellanic Cloud (LMC) offers a unique opportunity to resolve a stellar population in a starburst-like environment. Knowledge of the extinction towards this region is key for the accurate determination of stellar masses, and for the correct interpretation of observations of distant, unresolved starburst galaxies. 
 } 
 {Our aims are to construct an extinction law towards R136, and to measure the extinction towards individual sources inside the cluster. This will allow us to map the spatial distribution of the dust, to learn about dust properties, and to improve mass measurements of the very massive WNh stars inside the cluster. 
 } 
 {We obtain the near-infrared to ultraviolet extinction towards 50 stars in the core of R136, employing the `extinction without standards' method. To assure good fits over the full wavelength range, we combine and modify existing extinction laws.
 } 
 {We detect a strong spatial gradient in the extinction properties across the core of R136, coinciding with a gradient in density of cold gas that is part of an extension of the Stapler Nebula, a molecular cloud lying northeast of the cluster. In line with previous measurements of R136 and the 30 Doradus region, we obtain a high total-to-relative extinction ($R_V = 4.38 \pm 0.87$). However, the high values of $R_V$ are accompanied by relatively strong extinction in the ultraviolet, contrary to what is observed for Galactic sightlines.  
 } 
 {The relatively strong ultraviolet extinction towards R136 suggests that the properties of the dust towards R136 differ from those in the Milky Way. 
 For $R_{V} \sim 4.4$, about three times fewer ultraviolet photons can escape from the ambient dust environment relative to the canonical Galactic value of $R_{V} \sim 3.1$ at the same $A_{V}$. Therefore, if dust in the R136 star-bursting environment is characteristic for cosmologically distant star-bursting regions, the escape fraction of ultraviolet photons from such regions is overestimated by a factor of three relative to the standard Milky Way assumption for the total-to-selective extinction.
 Furthermore, a comparison with average curves tailored to other regions of the LMC shows that large differences in ultraviolet extinction exist within this galaxy. Further investigation is required in order to decipher  whether or not there is a relation between \RV and ultraviolet extinction in the LMC. 
  } 

   \authorrunning{Brands et al.}

   \maketitle


\section{Introduction \label{sec:introduction}}

The Tarantula Nebula (or 30 Doradus) in the Large Magellanic Cloud (LMC) is a vast, intrinsically bright star-forming region \citep[][]{1984ApJ...287..116K,2013AA...558A.134D}. 
With a large population of young, massive stars ($M \gtrsim 8$~\Msun) at a metallicity of about half Solar \citep[see][for an overview]{2007A&A...473..603M}, this region is reminiscent of giant starbursts observed in distant galaxies  \citep{2009MNRAS.399.1191C,2017Msngr.170...40C}. 
The proximity of 30 Doradus and its relatively unobscured nature allow us to resolve the stellar population of this starburst-like environment. This provides a unique opportunity to calibrate integrated quantities required for studies of starburst galaxies in which the stellar populations are unresolved \citep{1999IAUS..190..247T,2002ASPC..285..105B,2021ExA....51..887G}. 

At the heart of 30 Doradus lies the cluster R136, which hosts a rich population of (very) massive stars. 
In the centre of this young ($1-2$~Myr), massive ($M\sim2-10 \cdot 10^{4}~\mathrm{M}_\odot$; \citealt{1995ApJ...448..179H}; \citealt{2009ApJ...707.1347A}) cluster are the most massive stars observed to date \citep{1997ApJ...477..792D,2010MNRAS.408..731C,2020MNRAS.499.1918B,2022arXiv220211080B,2022ApJ...935..162K}. 
With its many (very) massive stars, R136 is responsible for $30\%-50\%$ of the overall ionising luminosity and wind power of 30 Doradus (\citealt{2019Galax...7...88C}; see also \citealt{2013AA...558A.134D}), and not surprisingly, the cluster plays an important role in stellar feedback mechanisms throughout this region \citep[e.g.][]{2011ApJ...738...34P,2019A&A...628A.113L,2021MNRAS.504.1627C}. 

In the current work, we focus on the extinction properties of sightlines towards R136. For the study of individual massive stars, knowledge of extinction is required in order to recover the intrinsic luminosity, which is an important factor in the determination of stellar masses; of particular interest are the masses of the three very massive WNh stars in the core of R136. 
Moreover, knowledge of the extinction provides insight into the dust properties of this starburst-like environment. This, in turn, might help us to understand the distant starburst galaxies, where stellar populations cannot be resolved and a proper characterisation of the dust properties is key for recovering the intrinsic ultraviolet (UV) brightness and identifying star formation \citep[e.g.][and references therein]{2020ARA&A..58..529S}. To properly account for the dust within (unresolved) starburst galaxies, usually an attenuation law is used rather than extinction law. Attenuation laws take into account not only the removal of light from the line of sight (extinction), but also the scattering of light into the line of sight. A widely used starburst attenuation law is that of \citet[][]{1994ApJ...429..582C,2000ApJ...533..682C}. For an in depth overview of dust attenuation laws in galaxies, we refer the reader to \citet{2020ARA&A..58..529S}. 

We investigate the dust properties by analysing the wavelength dependence of the extinction, often referred to as the `extinction law'. Extinction laws contain valuable information about the interstellar dust, as grain populations of different sizes are thought to account for the extinction at different wavelengths \citep{2003ARA&A..41..241D}. At long wavelengths, up to the optical range, extinction increases almost linearly and is associated with silicate and carbonaceous grains with sizes $> 250~\AA$. At shorter wavelengths, extinction due to these types of grains flattens, and the continuing increase of extinction in the UV and the peak around $2175$~\AA~are linked to smaller silicates (having sizes  $< 250$~\AA) and polycyclic aromatic hydrocarbon molecules \citep[PAHs; e.g.][]{2001ApJ...548..296W,2017ApJ...835..107X}. 
The relative fraction of each particle population determines the shape of the extinction curve. While its global shape is rather similar for all sightlines, indicating that the grains are rather uniformly mixed throughout the interstellar medium (ISM), more subtle differences between the curves exist. This shows that environmental factors can influence the grain populations. In this context, processes related to star formation are of particular importance. Examples of such processes are the erosion of molecular clouds under the influence of stellar winds and UV radiation, the shattering of grains by supernova explosions, and the coagulation of dust during cloud collapse \citep[e.g.][]{2018ARA&A..56..673G}. 

A parameter often used to characterise the shape of an extinction curve is the total-to-selective extinction, defined as $R_V \equiv A_V/E(B-V)$, with $E(B-V) = A_B - A_V$, and $A_V$ and $A_B$ being the extinction in the $V$ and $B$ photometric bands, respectively. The sightlines towards R136, and more generally 30 Doradus, display high values of $R_V$, with average values ranging from $R_V = 4.0$ to $R_V = 4.5$  \citep[][]{2013AA...558A.134D,2014AA...570A..38B,2014AA...564A..63M,2014MNRAS.445...93D,2016MNRAS.455.4373D,2016MNRAS.458..624C,2020MNRAS.499.1918B},  similar to sightlines towards star-forming regions in the Milky Way. Within 30 Doradus, large differences between  sightlines exist. For example, \citet{2014AA...564A..63M} find values in the range $R_V = 3.1-6.7$. Such variations in $R_V$ within a relatively small physical region are also observed for H~{\sc ii} regions in the Galaxy \citep{2018A&A...613A...9M}.  

$R_V$ is usually interpreted as a characterisation of the size distribution of dust grains, where sightlines with high values of $R_V$ are associated with a higher fraction of large dust particles, and a lower fraction of small dust particles. For the Milky Way, the dependence of UV extinction on $R_V$ has been investigated intensively,  consistently
revealing the same relation 
\citep[e.g.][]{1989ApJ...345..245C,1999PASP..111...63F,2019ApJ...886..108F}. While this might suggest that this behaviour is universal, we stress that this relation is empirical: a priori we would not expect a fixed relation between $R_V$, which is defined in a narrow optical region, and the extinction in the UV, which is attributed dominantly to different grain populations from those responsible for the extinction in the optical \citep[e.g.][]{2001ApJ...548..296W,2017ApJ...835..107X}. 

Analyses of extragalactic sightlines indeed show that the Galactic relation between $R_V$ and UV extinction is not universal. 
For example, \citet{1983MNRAS.203..301H} and \citet{1992ApJ...395..130P} derive average extinction curves for sight lines in the LMC corresponding to $R_V \approx 3.1$, the typical Galactic value, while they predict stronger extinction in the UV compared to Galactic laws of the same $R_V$. Also \citet{2003ApJ...594..279G}, who study two different sets of LMC sightlines, find curves that differ from the Galactic average. \citet{2019ApJ...878...31D}, who investigate three sight lines in 30 Doradus, find that the excess of large grains in this region (associated with the high $R_V$ values) does not seem to come at the expense of small grains. In other words, the high $R_V$ values found for R136 and 30 Doradus are not associated with UV extinction as weak as for Galactic sightlines with similarly high $R_V$. 
\citet{2019ApJ...878...31D} argue that supernova explosions have affected the dust population in the region.  

In this paper, we investigate the dust properties in 30 Doradus by studying 50 sightlines in the core of R136. While extinction laws in the near-infrared (NIR) and optical wavelength range of 30 Doradus and R136 have already been evaluated \citep[][]{2014MNRAS.445...93D,2014AA...564A..63M}, a detailed study of the UV extinction is lacking to date.   In the present paper, we measure extinction properties towards and around R136, that is, we construct a UV, optical, and NIR extinction law across the region.  This yields a spatial map of the extinction in and around the R136 cluster, an (average) extinction law, and insight into the dust properties. We use these results to obtain improved mass estimates of the WNh stars in the core of R136, the most massive stars known. 

The remainder of this paper is structured as follows. In \cref{sec:sample:data}, we describe our sample and data, and in \cref{sec:methods} we provide details of the 
methods used. In \cref{sec:results}, we present our main findings, which we discuss in a broader context in \cref{sec:discussion}. Finally, we summarise the main conclusions of our study in \cref{sec:conclusion}. 

\section{Sample and data \label{sec:sample:data}}

Our sample of the R136 core coincides with the sample of \citet{2022arXiv220211080B}, with the omission of six sources. We omit R136a6 because it comprises two sources that cannot be resolved in the UV spectroscopy, and R136a8, H49, H65, H129, and H162 because HST/WFC3 photometry of \citet[][see below]{2011ApJ...739...27D} is lacking for these sources.   
For all other stars, 50 in total, we compile a spectral energy distribution (SED) that spans from the UV to the NIR. In the UV, we use flux-calibrated spectroscopy, whereas for the optical and NIR we use photometry. 
An overview of the used observations can be found in \Cref{tab:data_overview_R136}; we describe the data in more detail below.  

\begin{table}[]
    \centering \small
    \caption{Data used to compile SEDs of 50 stars in the core of R136. }
    \label{tab:data_overview_R136}
    \begin{tabular}{l l l}
\hline \hline Regime & Instrument & Grating or filter  \\ \hline 
UV$^{a)}$  & HST/STIS & G140L ($1160-1710$~\AA) \\ 
Optical$^{b)}$  & HST/WFC3 & F336W, F438W, F555W, F814W$^{c,d)}$\\
NIR$^{e)}$  & VLT/SPHERE &  $J$  \\ 
NIR$^{f)}$  & VLT/SPHERE & $H$, $K_s$   \\ \hline 
\multicolumn{3}{p{8.5cm}}{\tiny}\\
\multicolumn{3}{p{8.5cm}}{\tiny $^{a)}$~\citet{2016MNRAS.458..624C}.  $^{b)}$~\citet{2011ApJ...739...27D}. $^{c)}$~These filters correspond roughly to the $U$, $B$, $V$ and $I$ bands, respectively. $^{d)}$~The stars H35,
H71, H73, H86, H116, H121, H135 lack an F814W magnitude. H69 lacks both an F555W and F814W magnitude. H139 lacks an F336W magnitude. For all other sources we have magnitudes for all specified bands. $^{e)}$~\citet{2017AA...602A..56K}.  $^{f)}$~\citet{2021MNRAS.503..292K}.}
\end{tabular}
\end{table}

\begin{figure}
    \centering
    \includegraphics[width=0.48\textwidth]{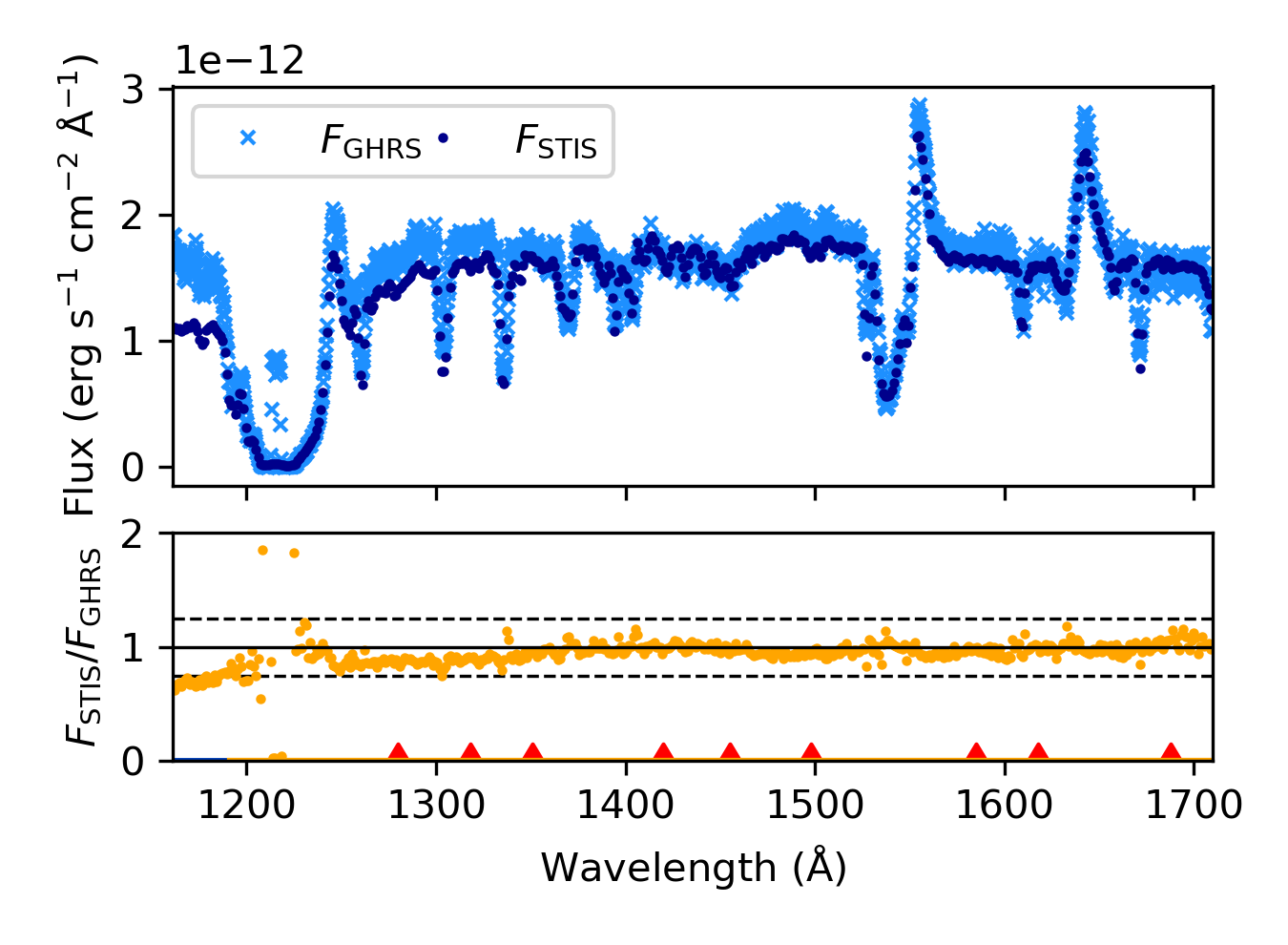}
    \caption{Integrated UV flux of the core of R136. The upper panel shows the GHRS spectrum of the inner 2$^\prime$ x 2$^\prime$ (\citealt[][]{1992ESOC...44..347H}; light blue crosses), as well as the summed STIS spectra used in this work (dark blue circles),  with both covering roughly the same inner region of the cluster. The bottom panel shows the ratio of the former two spectra (yellow dots) as well as the central wavelength of the synthetic UV bands used in this work (red triangles, see \cref{sec:obssed}). The dashed lines correspond to ratios of 0.75 and 1.25. \label{fig:integrated}}
\end{figure}

We use flux-calibrated HST/STIS spectroscopy presented in \citet{2016MNRAS.458..624C} for the far-UV part of the SED ($1150-1710$~\AA). These observations comprise 17 long-slit (52" x 0.2") contiguous pointings with grating G140L. In each slit, multiple sources are present (see \citealt{2016MNRAS.458..624C}, and \citealt{2022arXiv220211080B}, their Figs. 1). The spectra are extracted using {\sc multispec}, a package tailored to extracting spectra from crowded regions \citep{2005stisreport,2007multispec}. In the extraction process, the position of each source with respect to the slit centre is taken into account for accurate flux calibration. As a consequence of pointing inaccuracies, uncertainties in the calibration remain; we estimate these to be on the order of 10\% (see below), but in the most extreme cases they could be as large as a factor two. 
Small uncertainties in the flux calibration do not pose a problem for our analysis, as long as they are not systematic. 
To check our flux calibration for systematic uncertainties, we evaluate the integrated flux of the cluster core. To this end, we sum the flux-calibrated spectra of all stars in the cluster core. We then compare this integrated flux to the large aperture (2$^\prime$ x 2$^\prime$) HST/GHRS spectrum of the R136 core \citep{1992ESOC...44..347H}, which covers roughly the same area on the sky and for which the flux calibration is assumed to be reliable. Except for wavelengths $< 1200$~\AA, which are not included in our SED analysis, we find good agreement between the two. For all wavelengths considered in the fitting, the integrated fluxes match within 25\%, and for wavelengths $> 1350$~\AA~ the match is even within 10\% (\cref{fig:integrated}). On average, the sum of the STIS fluxes is $5\%$ lower than that of the GHRS spectrum. A modest difference is not surprising, as there are small discrepancies between the sky coverage and source extraction of the STIS and GHRS observations. 
All considered, we regard the calibration of the UV fluxes of \citet{2016MNRAS.458..624C} as reliable, and adopt the fluxes at face value. 

For the optical part of the SED, we use HST/WFC3 photometry of \citet{2011ApJ...739...27D}. We have magnitudes in the filters F336W, F438W, F555W, and F814W, which are roughly equivalent to the Johnson $U$, $B$, $V$, and $I$ filters, respectively. For eight stars, the F814W magnitude is missing, and for two stars the F336W or F555W magnitude is missing (see \cref{tab:data_overview_R136}). 
We also use the photometry of \citet{2011ApJ...739...27D} to estimate the extinction for 1657 stars in the outskirts of R136 (see \cref{res:surroundings}).  

The NIR photometry, complete for all stars in the sample, was taken in three bands ($H$, $J$, $K_s$) using VLT/SPHERE \citep{2017AA...602A..56K,2021MNRAS.503..292K}. We note that  $J$ and $K_s$ magnitudes are presented in \citet{2017AA...602A..56K}; and  $H$ and $K_s$ magnitudes in \citet{2021MNRAS.503..292K}. We adopt the  $H$ and $K_s$ magnitudes of the most recent analysis, and the $J$ magnitude of the first paper. In order to resolve the individual sources in the crowded R136 core, adaptive optics was employed, making absolute flux calibration challenging. \citet{2021MNRAS.503..292K} double check the calibration of their $H$ and $K_s$ magnitudes by comparing to the catalogue of \citet{2010MNRAS.405..421C} and report no systematic difference. \citet{2020MNRAS.499.1918B} carry out a similar check with the \citet{2017AA...602A..56K} catalogue, but only for the $K_s$ band, as a dataset for cross-checking the $J$ band does not exist. Nonetheless, we adopt the $J$ magnitude calibration and include these magnitudes in our analysis. This is because none of the checks on the absolute flux calibrations on the $H$ and $K_s$ bands give reason for concern, and the $J$ band calibration was carried out by \citet{2017AA...602A..56K} in the same manner as that of the $K_s$ band. We note that, in practice, the $J$ magnitudes have a negligible effect on the outcome of our analysis and removing them from the fitting would not affect our conclusions; this is because their uncertainties are large compared to those of the other bands, which minimises their weight in the fitting process. 

\section{Methods \label{sec:methods}}

In order to assess the extinction towards the sources in the core of R136, we employ the `extinction without standards' technique \citep[][]{1966ApJ...144..305W, 2005AJ....130.1127F}. 
This technique requires an observed SED, a model of the intrinsic SED, and an extinction law of which the shape can be parameterised. 
In addition, the process requires a fitting algorithm to find the optimal extinction curve parameters given the intrinsic and observed SED. 
We discuss each of these aspects below. 

For 1657 stars in the outskirts of R136 (all sources from the \citet{2011ApJ...739...27D} catalogue with $V<19$), we assess the extinction in a different way, namely by extrapolating the extinction properties we measure for the R136 core stars. To this end, we use the empirical relation between the colour $V-I$ and $A_V$ that we find for the core stars. We describe this process in \cref{res:surroundings}. 

\subsection{Observed SEDs\label{sec:obssed}}

\begin{figure}
    \includegraphics[width=0.46\textwidth]{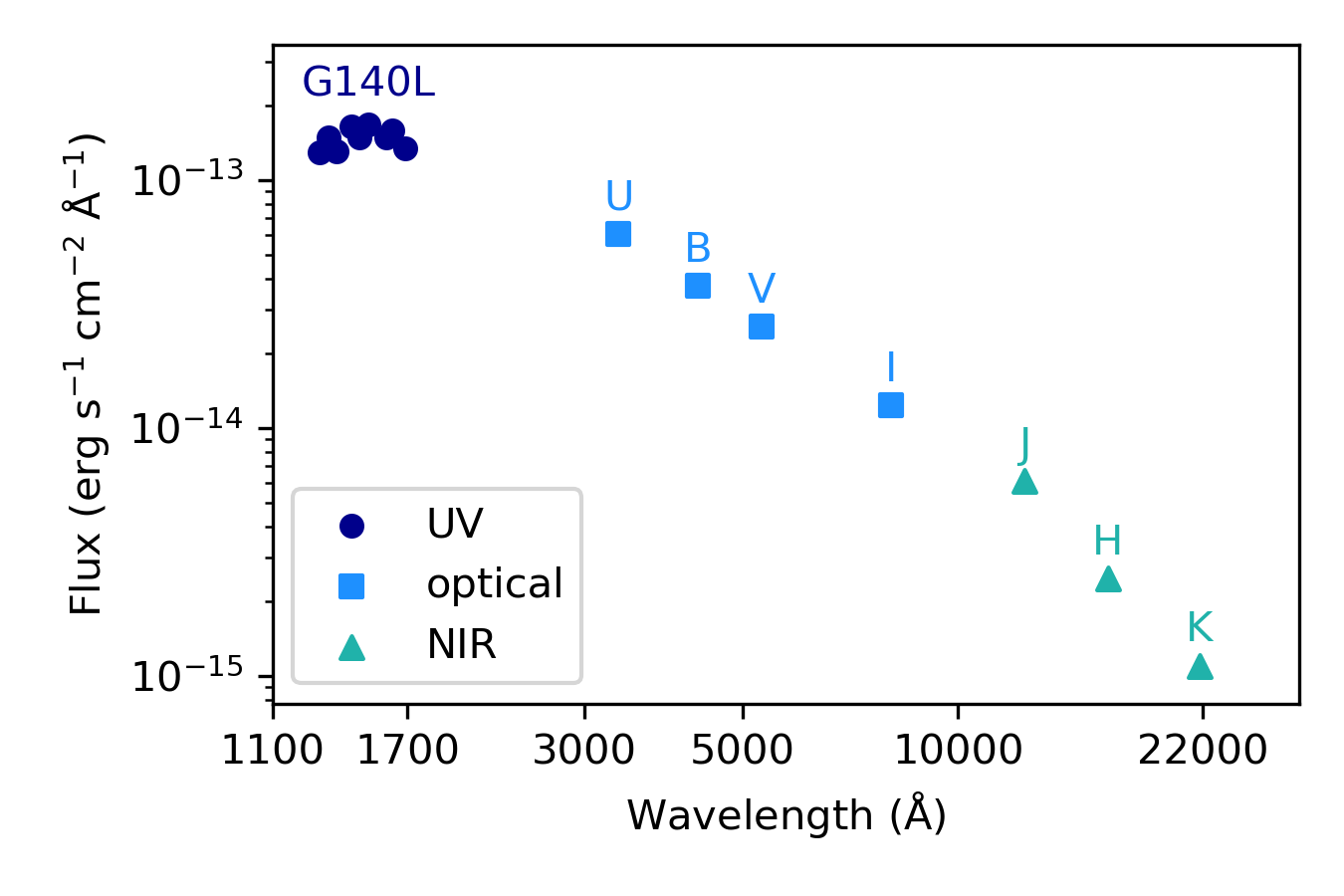}
    \caption{Example of an observed SED (source: R136a3) covering the UV (dark blue circles), optical (light blue squares), and NIR (turquoise triangles). The UV flux points are derived from flux-calibrated STIS spectroscopy (grating: G140L), and the optical and NIR fluxes come from broadband photometry; the (equivalent) name of each band is indicated.  
    \label{fig:example_SED}}
\end{figure}

We compile observed SEDs spanning from the UV to NIR (see also \cref{sec:sample:data}). The optical and NIR observed magnitudes 
are converted to fluxes using filter transmission curves taken from the SVO Filter Profile Service\footnote{ \url{http://svo2.cab.inta-csic.es/theory/fps/}}. 
For the UV, we bin the observed flux points by taking the flux average over nine different wavelength regions as specified in \cref{tab:app:wavesUV}. Taking such a flux average is equivalent to defining a filter with a transmission of 1.0 between the two wavelengths, and 0.0 outside that region. 
With the 4 optical and 3 NIR photometry points for the R136 core stars, the observed SEDs consist of 16 flux points in total. 

The photometric points from the UV to the NIR that we include in our fitting are spread over the SED in such a way that each point represents an approximately equal amount of energy radiated by the star. In other words, if we take an SED and integrate over the wavelength range spanned between the central wavelength of one filter and its neighbour, this amount of energy (per cm$^2$ per sec) is roughly equal for each filter. This ensures that the different parts of the SED are as equally represented as possible in the fitting process. We note that the results are robust to changes in the number of UV points that we include in our analysis; a change in the number of points has a negligible effect on the outcome of our fits and would not affect our conclusions. 
An example of one of our observed SEDs is shown in \cref{fig:example_SED}.

\subsection{Intrinsic SEDs \label{sec:intrinsic_seds}}

We compute models of intrinsic SEDs with the model atmosphere code {\sc Fastwind} \citep[][]{1997A&A...323..488S,2005A&A...435..669P,2012A&A...537A..79R,2016A&A...590A..88C, 2018A&A...619A..59S}. This code, which is tailored to hot stars with winds, solves radiative transfer subject to the solution of the non-local thermal equilibrium number-density rate equations and takes into account the effects of line blocking and line blanketing. Only a subset of the spectral lines is explicitly computed and the output SED therefore does not resolve individual spectral lines. For our analysis, this is not directly an issue, as we are not fitting individual lines. Still, the absence of lines in the model SED can result in small differences in flux of the integrated bands. However, the effect of this on the outcome of our analysis is only minor. 
We mimic the magnitude of the effect by representing the UV absorption-line forest by an overall decrease in the UV model flux of 10\%.  In this case, the best-fitting UV extinction is only slightly lower ($\Delta c_{2} = -0.05$) compared to the case where the continuum is unmodified. This change is within the typical range for uncertainties of individual $c_{2}$ values. Moreover, the best-fit values for $A_{5495}$, $R_{5495}$, and luminosity remain unaffected. We conclude that the effect of using SEDs without explicit spectral lines is only minor. 

To further ensure that our models are valid as intrinsic SEDs, we compare {\sc Fastwind} models with low mass-loss rates ($10^{-10}$~\Msun/yr) to hydrostatic models from the TLUSTY grid of \citet{2003ApJS..146..417L}, which have been used for other extinction studies (e.g. \maiza). First, we assess how the {\sc Fastwind} models behave in the $U-B$ versus $B-V$ plane to make sure that the behaviour around the Balmer jump is correct, as the latter is only modelled in a crude way in the {\sc Fastwind} SEDs. We find that the qualitative behaviour of the two sets of models is similar, and that the absolute differences between the {\sc Fastwind} and TLUSTY colours are never more than 2\%. Second, we compare the behaviour in the $J-H$ versus $H-K$ plane of the TLUSTY and {\sc Fastwind} SEDs. Again, we find that the qualitative behaviour of the two sets of models is similar, and furthermore, that the differences are never larger than 1\%. Overall, we conclude that the {\sc Fastwind} models are suitable to be used as intrinsic SEDs for this study. 

For the stellar parameters, we adopt the values of the optical and UV analysis of \citet{2022arXiv220211080B} for stars in the core of R136. 
We note that for computing these models, we need to adopt a stellar radius $R_*$, while this is one of the parameters that we constrain in this study (see \cref{sec:fitting}). 
However, for small changes in $R_*$, the structure of the atmosphere will not change significantly and this means that we can change $R_*$ of the model simply by multiplying the output SED with a factor that represents an increase or decrease in the stellar surface area. Using this approach, we need only one model SED per star, and are still able to fit $R_*$. 

\subsection{Extinction law\label{sec:extcurves}}

\begin{figure}
    \centering
    \includegraphics[width=0.48\textwidth]{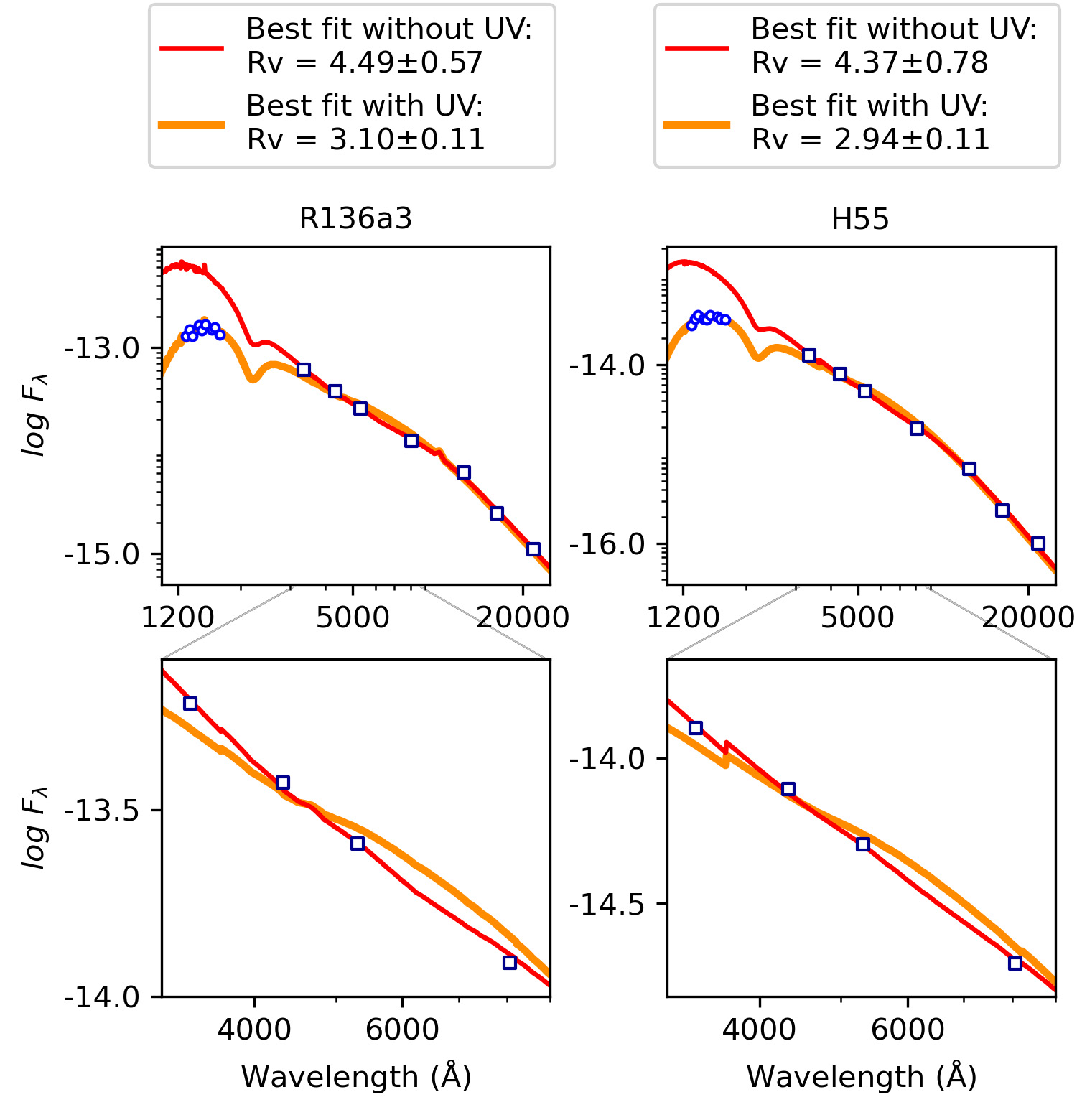}
    \caption{Example fits of SEDs of the stars R136a3 (left; WNh) and H55 (right; O2 V((f*))z), adopting the standard Galactic law of \fitz. The figure shows that this law is not suitable for our analysis: the Galactic dependence of UV extinction on $R_V$ does not hold for sightlines towards R136. 
    In other words, the reddened flux cannot be modelled accurately over the full wavelength range: 
    an $R_V$ based on the optical and NIR photometry gives a total mismatch with the observed UV fluxes, whereas a much lower value ---that can reproduce the UV fluxes--- gives a very poor fit to the observed slope in the optical. 
    The top panels display the flux from UV to NIR; the bottom panels zoom in on the optical regime for which UBVI photometry is available. Circles and squares indicate observed fluxes in the UV, and optical or NIR, respectively. 
    The best-fitting reddened model including the UV is shown in orange; the best fit excluding the UV constraints in red. For this example, we used the law of \fitz, but the same behaviour is seen with other $R_V$-dependent Galactic laws.  
    \label{fig:nogalacticRvbehaviour}}
\end{figure}

When considering the SED from the NIR to the UV, none of the existing extinction laws seem suitable for sightlines towards 30 Doradus. The average LMC curves of \citet{1983MNRAS.203..301H} and \citet{1992ApJ...395..130P} correspond to an optical slope of $R_V \approx 3.1$, whereas higher values in the range of $R_V = 4.0-4.5$ are found  for 30 Doradus (for references, see \cref{sec:introduction}). Also, the average LMC curve of \citet{2003ApJ...594..279G}, who find $R_V = 3.41$, does not match the value of 30 Doradus, nor does their curve towards LMC2 (more commonly referred to as `LMC SGS~2'), lying southeast of 30 Doradus. For LMC SGS~2, these latter authors find a rather low value of $R_V = 2.76$. \citet[][]{1999ApJ...515..128M} also find low values of $R_V$, both for stars in LMC SGS~2 ($R_V \leq 3.31$) and for stars elsewhere in the LMC ($R_V \leq 2.61$)\footnote{A note on nomenclature: \citet[][]{1999ApJ...515..128M} use `30 Doradus' to refer to a much larger region than we do. In this study, we use the name 30 Doradus to refer to the inner $\approx 10'$ as shown in for example Fig. 2 of \citet{1991IAUS..148..145W} and Fig. 1 of \citet{2020Msngr.181...22E}, whereas \citet[][]{1999ApJ...515..128M}, but also for example \citet{1985ApJ...288..558C}, consider a larger region when they refer to 30 Doradus; sometimes these authors use 30 Doradus interchangeably with LMC SGS~2.}. The law of \citet{2014AA...564A..63M} is tailored to 30 Doradus, but is only valid in the optical and NIR. 

Also, the Galactic laws, many of which allow for a varying $R_V$, are not suitable for 30 Doradus when the analysis is extended to the UV. We demonstrate this in \cref{fig:nogalacticRvbehaviour}, by showing the fits of two stars, with and without UV, while adopting the Galactic law of \citet[][hereafter \fitz]{1999PASP..111...63F}, which has a typical Galactic $R_V$-dependent UV extinction. The two stars shown are R136a3 (WNh) and HSH95-55 (or H55, O2 V((f*))z), representative of components within R136. When fitting only the NIR and optical, we reproduce the high values of \RV obtained by previous studies. For example, for R136a3, we find $R_V = 4.49 \pm 0.57,$ and for H55, we obtain $R_V = 4.37 \pm 0.78$. Indeed, the lower panels of the figure clearly show that a high \RV better reproduces the observed optical photometry. However, the high values of \RV very poorly reproduce the UV part of the SED. To fit the full SED, that is, including the UV, lower values are needed: $R_V = 3.10 \pm 0.11$ for R136a3 and $R_V = 2.94 \pm 0.11$ for H55. 
The behaviour shown in \cref{fig:nogalacticRvbehaviour} for R136a3 and H55 is observed for all stars in the R136 core, and is also seen when adopting a different Galactic law, such as that of \citet{1989ApJ...345..245C} or that of \citet{2019ApJ...886..108F}. 

As no existing extinction law meets our requirements, we need to either adjust an existing extinction law, or derive our own. The data we have available are insufficient to derive a completely new law: in the NIR and optical, we have only a few data points, and in the near-UV (between the $U$ band and the red end of the STIS/G140M grating at 1710~\AA), we lack data altogether. Given the complex shape of extinction curves, the number of free parameters we would need to fit would exceed the number of data points we have in these wavelength regions. We therefore decided to modify an existing law. 

The parameterisation of our law is similar to that of \fitz. These latter authors provide an $R_V$-dependent Galactic law consisting of a cubic spline going through anchor points at fixed wavelengths for the optical and NIR, and a UV part ($\lambda \lesssim 3000$~\AA) that is described by the simple parameterisation of \citet{1990ApJS...72..163F}. The strength of extinction at each optical and NIR spline point is tailored to Galactic sightlines and is \RV-dependent. For this work, we do not adopt the optical and NIR spline point values of \fitz, but rather use the shape of the extinction law of \maiza.
\maiza provide a family of extinction laws that depend on $R_{5495} \equiv A_{5495}/(A_{4405} - A_{5495})$,  the monochromatic equivalent of \RV. We therefore use monochromatic quantities throughout the paper, with the exception of \cref{sec:dis:RV}, where we compare with other studies that adopted broadband quantities. We note that since the extinction towards R136 is moderate, and the SEDs of the stars we study are fairly similar, the differences between the broadband quantities ($A_V$ and $R_V$) and their monochromatic equivalents ($A_{5495}$ and \Rmono) are small. We refer the reader to \citealt{2013hsa7.conf..583M}, for a discussion on monochromatic versus broadband quantities. 

\maiza use a combination of the seventh-order polynomials of \citet{1989ApJ...345..245C} and correction factors on those polynomials to express their law. We convert this functional form to the format of \fitz, that is, we use a linear relation for expressing the value of each spline point\footnote{For one spline point, \fitz use a quadratic function; we use linear functions in all cases.}, so that the dependence on \Rmono of each spline point is explicitly provided by a formula for the sake of clarity. 
Furthermore, we omit the spline points of \maiza for $x\geq2.7~\mu$m$^{-1}$ ($\lambda \geq 3703$~\AA); for these wavelengths we adopt the UV prescription of \citet[][see below]{1990ApJS...72..163F}. We add an extra spline point at $x = 3.0~\mu$m$^{-1}$ ($\lambda = 3304$~\AA), and the $R_{5495}$-dependent extinction at this point we take from the law of \maiza. 
The latter spline point is crucial in order to fit \Rmono, reflecting the slope of the extinction curve in the optical and NIR, and the slope of the UV extinction as parameterised by \citet{1990ApJS...72..163F}, as truly independent parameters. All adopted spline points are listed in \cref{tab:newsplinesMA14}. 
Our parameterisation matches the law of \maiza exactly at the spline points, and within $\leq 0.1~\%$ for all other optical and NIR wavelengths. 

Before we continue to describe the UV part of our law, we note that the slope of the NIR part of the extinction law of \maiza is possibly not ideal for extinction towards R136. The NIR part of the \maiza law can be approximated by a power law of the form $A_\lambda \propto \lambda^{\alpha_\mathrm{NIR}}$ with $\alpha_\mathrm{NIR} = 1.61$; this value is lower than values obtained by recent studies of Galactic extinction, which find $\alpha_\mathrm{NIR} = 2.1-2.4$ \citep{2009MNRAS.400..731S,2019A&A...630L...3N,2020MNRAS.496.4951M}. Upon inspecting the residuals of our fits as a function of $A_V$, we see clear patterns for all NIR bands, indicating that the law is not completely adequate. However, we adopted it as it is because we have only a few data points in the NIR, and it is beyond the scope of this paper to improve the NIR extinction curve. 

\begin{table}[]
    \caption{\Rmono-dependent values of optical and NIR spline anchor points based on the law of \maiza. 
    \label{tab:newsplinesMA14}}
    \centering
    \begin{tabular}{r l l}
    \hline\hline $\lambda$ (\AA) & $\lambda^{-1}$ ($\mu$m$^{-1}$) & $A_\lambda/(A_{4405}-A_{5495}$) \\ 
    \hline 
$\infty$ & 0.000 & 0.0 \\
26500 & 0.377 & $-0.1097 + 0.1195 \ \times$ \Rmono \\
18000 & 0.556 & $-0.2046 + 0.2228 \ \times$ \Rmono \\
12200 & 0.820 & $-0.3826 + 0.4167 \ \times$ \Rmono \\
10000 & 1.000 & $-0.5270 + 0.5740 \ \times$ \Rmono \\
8696  & 1.150 & $-0.6392 + 0.7147 \ \times$ \Rmono \\
5495  & 1.820 & $-0.0002 + 1.0000 \ \times$ \Rmono \\
4670  & 2.141 & $\phantom{-}0.7455 + 1.0023 \ \times$ \Rmono \\
4405  & 2.270 & $\phantom{-}1.0004 + 1.0000 \ \times$ \Rmono \\
4110  & 2.433 & $\phantom{-}1.3149 + 0.9887 \ \times$ \Rmono \\
3704  & 2.700 & $\phantom{-}1.7931 + 0.9661 \ \times$ \Rmono \\
3304  & 3.027 & $\phantom{-}2.2580 + 0.9689 \ \times$ \Rmono \\ \hline 
\multicolumn{3}{p{7cm}}{\footnotesize \textit{Notes.} For the interpolation between these points we use a cubic spline and the function \texttt{interpolate.splrep} of the Python package \texttt{scipy} \citep{2020SciPy-NMeth}. \Rmono is the monochromatic equivalent of \RV: $R_{5495} \equiv A_{5495}/(A_{4405} - A_{5495})$.}
      \end{tabular}
\end{table}

For the UV part of the curve, we use the parameterisation of \citet{1990ApJS...72..163F}. 
Expressed in terms of the quantity $k(\lambda - V)$ $\equiv (A_\lambda - A_V)/(A_B - A_V) \equiv E(\lambda - V)/E(B-V)$, the UV part of the law of \fitz has the following form: 
\begin{equation} \label{eq:uvfunc}
    \frac{E(\lambda - V)}{E(B-V)} =\\ \left\{ 
\begin{array}{ll}
c_1 + c_2 x + c_3 D(x,x_0,\gamma)                  &  x \leq c_5     \\
c_1 + c_2 x + c_3 D(x,x_0,\gamma) +  c_4 a_1 (x-c_5)^2   &    \;\;  \\
             \hspace{2.5cm} + c_4 a_2 (x-c_5)^3 &  x > c_5,  \;\; 
\end{array}
\right. 
\end{equation}
where $x\equiv 1/\lambda$~$\mu$m$^{-1}$, $a_1 = 0.5392,$ and $a_2 = 0.05644$\footnote{As we express the optical and NIR part of the law in monochromatic quantities, following \maiza, the $V$ and $B$ band quantities in \cref{eq:uvfunc} are replaced by monochromatic values at $\lambda = 5495$~\AA~ and $\lambda = 4405$~\AA, respectively.}. 
The parameters $c_1$ and $c_2$ relate to a linear background (see below); $c_3$ indicates the strength of the \bump (see below), and $c_4$ and $c_5$ indicate the strength and start of the far-UV curvature, respectively. \citet{1990ApJS...72..163F} and \fitz do not treat the start of the UV curvature as a free parameter, but adopt a fixed value of $c_5 = 5.9$; we do the same. 
We leave $c_4$ as a free parameter, so that we can constrain the UV curvature in the extinction curve of sightlines towards R136. 
The \bump is described by the Drude profile:
\begin{equation}
\label{eq:drude}
D(x,x_0,\gamma) = \frac{x^2}{(x^2-x_0^2)^2 +x^2\gamma^2},
\end{equation}
with $x_0$ being the position and $\gamma$ the width of the bump. We do not fit the data of the bump as we lack observations here, and instead adopt the values that \citet{2003ApJ...594..279G} find for LMC SGS~2 (near 30 Doradus), that is, $c_3 = 1.463$, $\gamma = 0.945,$ and $x_0 = 4.558$. 

Lastly, and key for our study, we discuss the parameters $c_1$ and $c_2$. These parameters do not have a fixed value in the average curve of \fitz, but instead are expressed in terms of \RV and each other. For Galactic sightlines, \fitz derive: 
\begin{equation}\label{eq:fitz_uv_orig}
\begin{split}
        c_2 &= -0.824 + 4.717/R_V  \\
        c_1 &= \phantom{-}2.030 - 3.007c_2.
\end{split}
 \end{equation}
It is the dependence on \RV that links the shape of the extinction in the optical and the NIR to that in the UV. However, while the relation in \cref{eq:fitz_uv_orig} is typical for Galactic sightlines, it does not apply to our case (see \cref{fig:nogalacticRvbehaviour}). Therefore, in the present study, we leave $c_2$ as a free parameter in the fitting process. In other words, we remove the Galactic dependence of the UV extinction on \RV from our law: the strength of extinction in the UV ($c_2$) and the slope of extinction in the optical (\Rmono) will be fitted independently. While we fit the slope of the UV extinction, the equation for $c_1$ we leave unchanged. This is because $c_2$ and $c_1$ are (to some extent) degenerate, and we do not have enough data to break this degeneracy. Experiments where we leave $c_1$ free in the fitting process show that, indeed, it is not possible to disentangle these two parameters with our data. 

Briefly, we use the optical and NIR law of \citet{2014AA...564A..63M}, which we transform to the functional form of \citet{1999PASP..111...63F}. For the UV, we adopt the parameterisation of \fitz and \citet{1990ApJS...72..163F}. The \bump is described as in \citet{2003ApJ...594..279G}, and the parameters \Rmono (slope in the optical; monochromatic equivalent of \RV), $c_2$ (strength of UV extinction), and $c_4$ (far-UV curvature) are free parameters. 

\begin{figure*}
    \centering
    \includegraphics[width=0.98\textwidth]{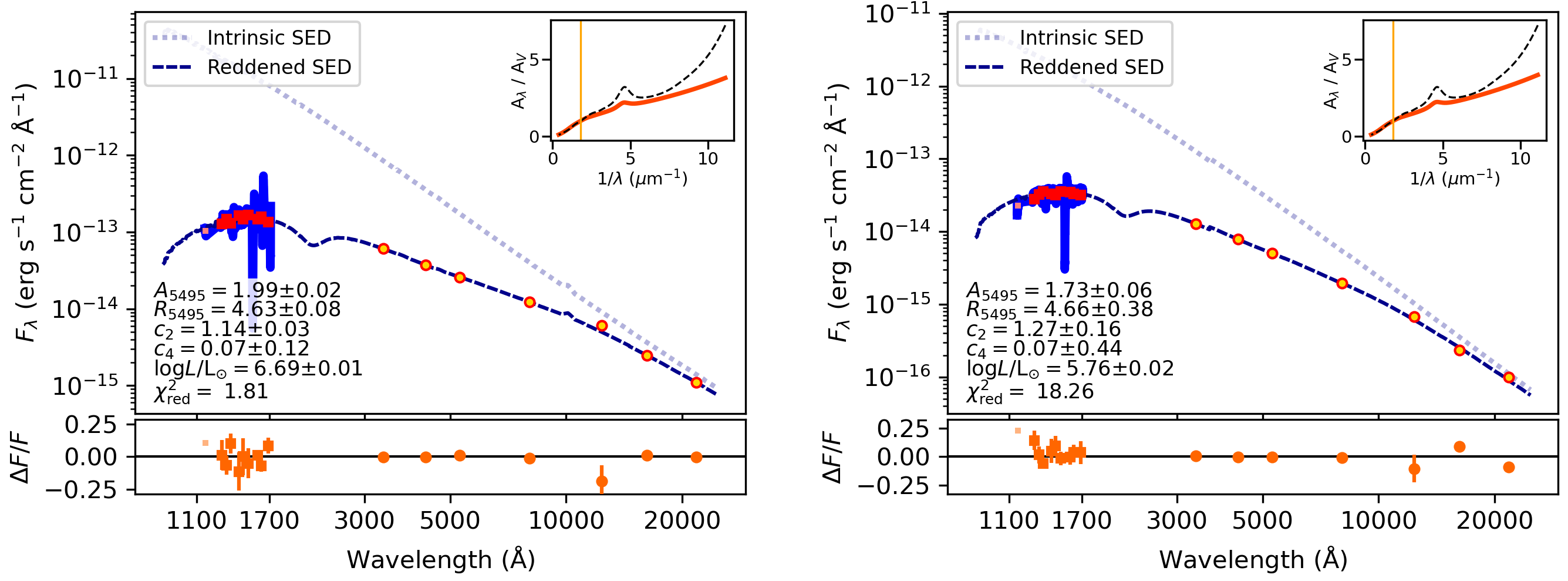}
    \caption{Example fits of the stars R136a3 (WN5h; left) and H55 (O2 V((f*))z; right).  
    In the upper part of each panel, we show the photometric points that were used for the fitting;  with red solid squares and yellow filled circles indicating points in the UV and optical or NIR, respectively.  
    The thick solid blue line in the background shows the UV spectroscopy from which the UV photometric points were derived. The light blue dotted and dark blue dashed lines in the top panels show the adopted intrinsic SED, and the reddened SED (resulting from the fitting process), respectively. The lower panel shows residuals, and the best-fit parameters are printed in the upper panel. 
    Lastly, each panel contains an inset where the shape of the extinction curve under consideration (solid orange line) is compared to the Galactic curve of \citet[][$R_V = 3.1$; black dashed line]{1989ApJ...345..245C}.  \label{fig:Example_fits}}
    
\end{figure*}

\subsection{Fitting \label{sec:fitting}}

For each individual star, we optimise the free parameters of the adopted extinction law (\cref{sec:extcurves}) in order to find the best match between the observed SED and the reddened model SED. 
We fit five parameters: \Rmono, affecting the shape of the optical and NIR part of the curve; $c_2$ and $c_4$, relating to the UV part of the curve; the extinction in the $V$ band $A_V$; and the stellar angular radius $\theta_R = R_*/d$, where $R_*$ is the physical radius of the star, and $d$ is the distance. We adopt a fixed LMC distance of $d = 49.59$~kpc for all stars \citep{2019Natur.567..200P} and therefore in practice the parameter that we vary is $R_*$, or,  equivalently, the stellar luminosity $L_*$. We assume values for $T_{\rm{eff}}$ as described in  \cref{sec:intrinsic_seds}. 
We stress that changes in the parameters $A_V$ and $R_*$ have different effects and are not degenerate, as changing $A_V$ affects the extinction at each wavelength differently\footnote{Where the change as a function of wavelength is dictated by the shape of the adopted extinction curve.}, while changing $R_*$ changes the flux of the intrinsic SED by an equal factor for all wavelengths. 

In order to find the best-fitting values of the free parameters $A_V$, $R_*$, and those describing the shape of the curve (\Rmono, $c_2$, and $c_4$), we use the \texttt{minimize}  function of the Python package \texttt{lmfit}\footnote{We used \texttt{lmfit} version 1.0.3 in combination with Python version 3.8.5. Furthermore, we made use of \texttt{numpy} version 1.19.2 \citep{harris2020array} and \texttt{scipy} version 1.8.1 \citep{2020SciPy-NMeth}.} \citep[][]{2014zndo.....11813N}. We adopt the least-squares Levenberg-Marquardt method for the minimisation, where we minimise the $\chi^2$ value:
\begin{equation}
    \chi^2= \sum_{i=0}^N \left( \frac{\mathcal{F}_{{\mathrm{mod}},i} - \mathcal{F}_{{\mathrm{obs}},i}}{ \mathcal{O}_{i}} \right)^2, 
\end{equation}
where $N$ is the number of data points $i$ of the SED that is considered in the fit, $\mathcal{F}_{{\mathrm{mod}},i}$  the reddened flux of the model, $\mathcal{F}_{{\mathrm{obs}},i}$  the observed flux of the reddened star, and 
$\mathcal{O}_{i}$ is the observational uncertainty on each flux point. For the optical and NIR, we obtain observational uncertainties from the literature; for the UV, we use the standard deviation of the fluxes in each synthetic band as an uncertainty (\cref{sec:obssed}). The reduced-$\chi^2$, the $\chi^2$ value per degree of freedom, is defined as  $\chi^2_\mathrm{red} \equiv \chi^2/n_\mathrm{dof}$, with $n_\mathrm{dof} = N - n_\mathrm{free}$ the degrees of freedom, where $n_\mathrm{free}$ is the number of free parameters. 

In order to obtain $\mathcal{F}_{\mathrm{mod},i}$, we first obtain the  distance-corrected reddened model flux, $f_\lambda$, by applying the extinction law under consideration to the intrinsic model spectrum: 
\begin{equation}
f_{\lambda} = \theta^2_R F_{\lambda}10^{-0.4 A_{\lambda}},
\end{equation} 
with $F_{\lambda}$ the intrinsic (unreddened), distance-corrected model flux, and $A_\lambda$ the extinction in magnitudes as a function of wavelength, dictated by the adopted law and $A_V$. As discussed above, $\theta_R$ scales with stellar radius (luminosity) and is a free parameter, as is $A_V$. After obtaining the reddened model SED, we get the fluxes $\mathcal{F}_{\mathrm{mod},i}$, by using the transmission curves from filters used for the construction of our observed SEDs. 

\section{Results \label{sec:results}}

\subsection{An \Rmono-dependent average extinction law for R136 \label{sec:results:newcurve}}

\begin{figure}
    \centering
    \includegraphics[width=0.30\textwidth]{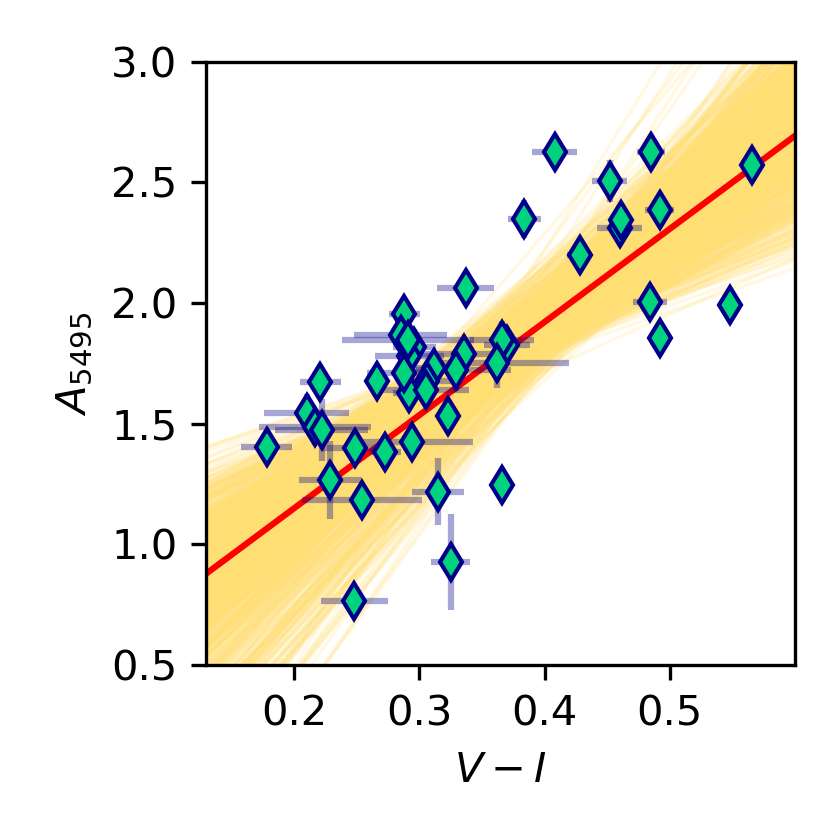}
    \caption{Extinction $A_{5495}$ versus observed colour $V-I$. Each diamond corresponds to a star in the core of R136; light-blue error bars indicate $1\sigma$ uncertainties. The red solid line is the best fit through all points, and in yellow we show the bootstrapped uncertainties on the linear fits, represented by linear fits to 1000 randomly chosen samples.
    \label{fig:c2_vs_Rv}}
\end{figure}

\begin{figure}
    \centering
    \includegraphics[width=0.48\textwidth]{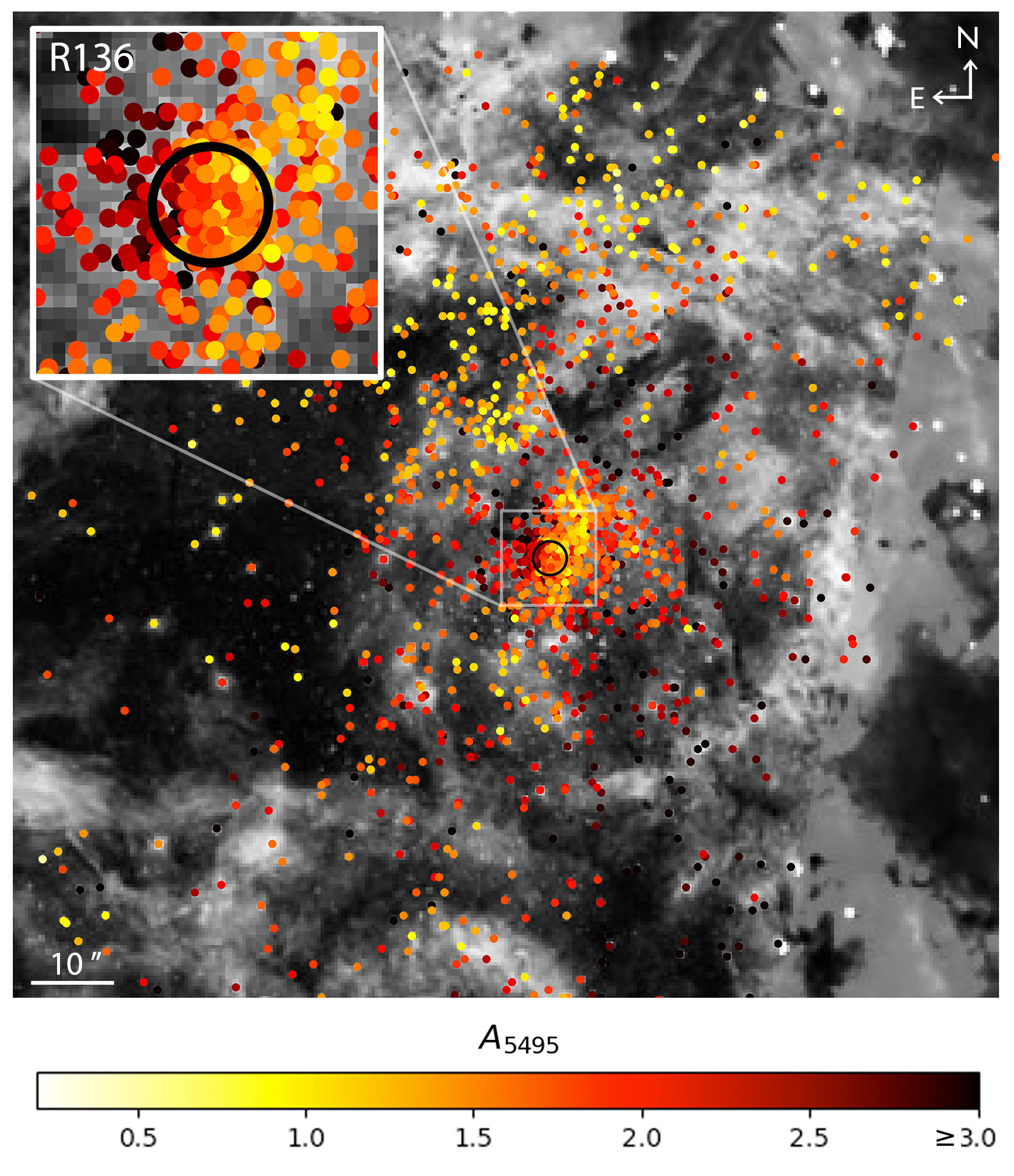}
    \caption{Extinction map of the R136 cluster and surroundings. 
  The dots indicate stars of the \citet{2011ApJ...739...27D} sample for which $A_V$ was estimated using $V-I$ (see \cref{res:surroundings}), as well as stars in the core of the R136 cluster, for which $A_V$ was determined with a full SED fit. 
  The core of the cluster is indicated with a black circle, and a zoom onto the core of the R136 cluster is shown in the inset in the upper left corner. We note that in order to make more details visible, the dots in the inset plot are of a smaller size relative to the background image than those of the main panel. 
    Background image credits: ESO/R. Fosbury (ST-ECF), R. O'Connell (University of Virginia, Charlottesville), and the Wide Field Camera 3 Science Oversight Committee (colours adapted). 
    \label{fig:30dor_vfts_demarchi}}
\end{figure}

We fit the SEDs of the stars in the core of R136 and for each star we obtain best-fit values and uncertainties for \Amono, \Rmono, $c_2$, $c_4$, and luminosity. We find a large range of extinction values throughout the cluster, with $A_{5495}$ ranging from $A_{5495} = 0.76\pm0.06$ (H108) to $A_{5495} = 2.63\pm 0.05$ (H36), and \Rmono ranging from $R_{5495} = 1.91\pm0.26$ (H108) to $R_{5495} = 5.79\pm0.31$ (H36), and cluster averages of $A_{5495} = 1.71\pm0.41$ and of $R_V = 4.38 \pm 0.87$. The best-fit parameters of individual stars are listed in \cref{tab:app:R136results}. Two example fits are shown in \cref{fig:Example_fits}. 

The $\chi^2_\mathrm{red}$ of our best fits range from $\chi^2_\mathrm{red} = 0.52$ (for H90) to $\chi^2_\mathrm{red} = 164.30$ (for H120), with an average of $\chi^2_\mathrm{red} =  18.3$ and a median of $\chi^2_\mathrm{red} = 9.6$. These values are relatively high, suggesting that the adopted extinction law does not describe the dust properties sufficiently well, and/or that the adopted uncertainties are underestimated. The $H$ and $K_s$ bands dominate the high $\chi^2$ values: relative to the observational uncertainties the residuals in $H$ and $K_s$ are on average six times larger than in the other bands. As noted before, the extinction law in the NIR might be imperfect. However, the data that we have do not allow us to improve on this and we therefore accept the values of our fit. 

We find a strong gradient in extinction $A_{5495}$ across the core of R136. We discuss this in more detail in \cref{sec:res:spatial} and Section \ref{sec:dis:spatial}; 
for now it is important to note that the wide range of values in \Rmono that we find throughout the core of R136 correspond to similar values of $c_2$, for which we find values in the range of $c_2 = 0.78-2.36$, with an average of $c_2 = 1.30\pm0.22$. 
The values show no significant trend as a function of \Rmono, contrary to what is observed for Galactic sightlines (\cref{eq:fitz_uv_orig}). 

\begin{figure*}
    \centering
    \includegraphics[width=0.85\textwidth]{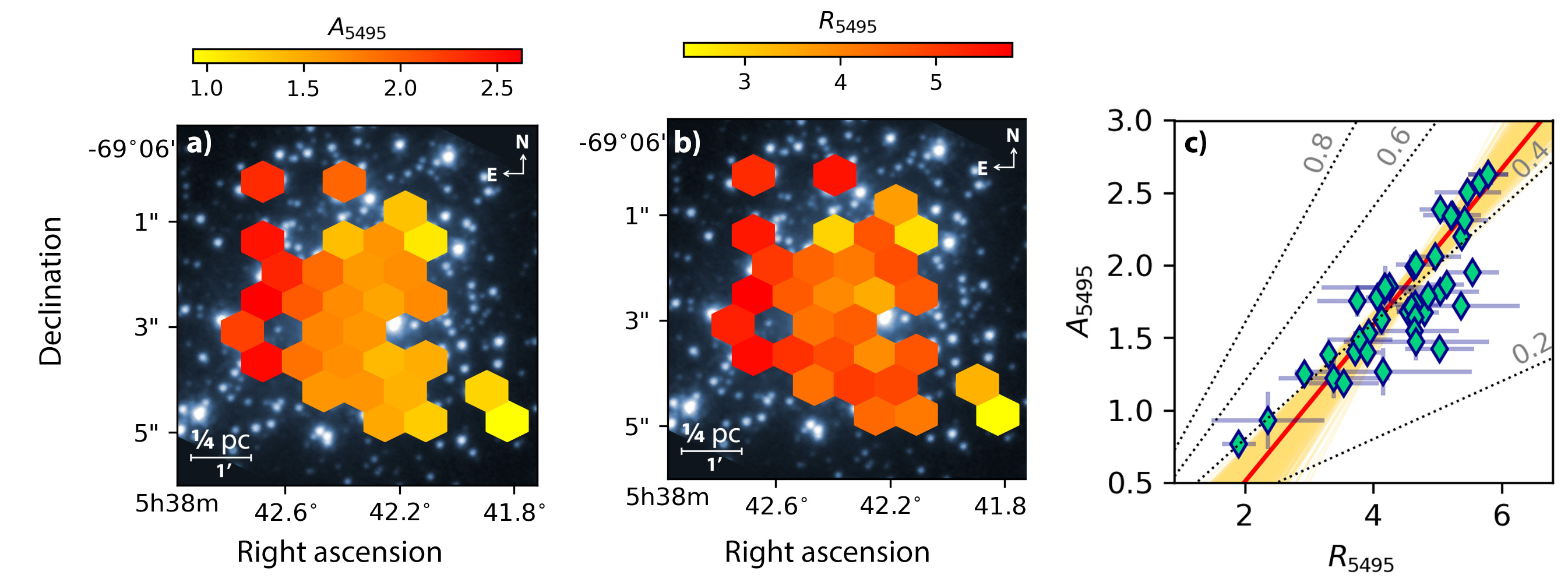}
    \caption{Spatial trends in $A_V$ and \Rmono in the core of R136. Panels (a) and (b) show maps of $A_{5495}$ and \Rmono overplotted on HST/WFC3 V-band (F555W) photometry \citep{2010AAS...21522205O}. The colour of each hexagon corresponds to the average value of the quantity in that region of the core, with red tints corresponding to higher values of $A_{5495}$ and \Rmono  and the yellow tints to lower values. A clear spatial trend is visible for $A_{5495}$, with higher values of $A_{5495}$ towards the east of the cluster (see also \cref{fig:spatial_trends_fit}). A similar, albeit weaker trend is visible for \Rmono. Panel (c) shows that there is a strong correlation between $A_{5495}$ and \Rmono (colours as in \cref{fig:c2_vs_Rv}). The dotted lines in the background indicate constant values of $A_{4405}-A_{5495}$ ranging from 0.2 to 0.8; the obtained relation implies $A_{4405}-A_{5495} \approx 0.4$.}
    \label{fig:2maps}
\end{figure*}

We can now construct an average \Rmono-dependent extinction law towards R136. We do this by adopting the  \Rmono-dependent optical and NIR law \maiza (using the spline point parameterisation of \fitz, as described in \cref{sec:extcurves}), and adopting the UV parameterisation of \citet{1990ApJS...72..163F}, assuming average values of $c_2$ and $c_4$ that we derive from the fitting, and \bump parameters of \citet{2003ApJ...594..279G}. 
We discuss how this curve compares to other LMC extinction curves in \cref{sec:dis:RV}. The parameters of the extinction curve towards R136 are summarised in \cref{tab:the_R136_curve}; a Python implementation of the law is presented in \cref{app:python_equation}, and can also be found on Github\footnote{\label{githubfootnote} \url{https://github.com/sarahbrands/ExtinctionR136/}}.

\begin{table}
    \centering
    \caption{Average \Rmono-dependent extinction law for R136.}
     \small 
    \begin{tabular}{p{1.7cm} p{2.1cm} p{3.8cm}}
   
    \hline \hline 
    Parameter & Value(s) & Source \\\hline 
    Spline points & See \cref{tab:newsplinesMA14} & \maiza, \fitz, this work  \\
    $c_2$ & $1.30$ & This work \\
    $c_1$ & 2.030 -- 3.007 $c_2$ & \fitz \\
    $c_3$ & 1.463 & \citet{2003ApJ...594..279G}\\
    $c_4$ & $0.09$ & This work\\
    $c_5$ & 5.9 & \citet{1990ApJS...72..163F}\\
    $x_0$ & 4.558 & \citet{2003ApJ...594..279G}\\
    $\gamma$ & 0.945 & \citet{2003ApJ...594..279G}\\\hline 
    \multicolumn{3}{p{7cm}}{\tiny} \\
    \multicolumn{3}{p{8.4cm}}{\small \textit{Notes.} Only the optical/NIR part of the law has a dependence on \Rmono. The optical/NIR part of the law is as derived by \maiza, but we adopt a functional form similar to that of \fitz; see \cref{sec:extcurves}. A Python implementation of the full law can be found in \cref{app:python_equation} and on Github$^{\ref{githubfootnote}}$.} \\
    \end{tabular}
    \label{tab:the_R136_curve}
\end{table}

\subsection{Extinction towards the outskirts of R136 \label{res:surroundings}}

\begin{figure}
    \centering
    \includegraphics[width=0.50\textwidth]{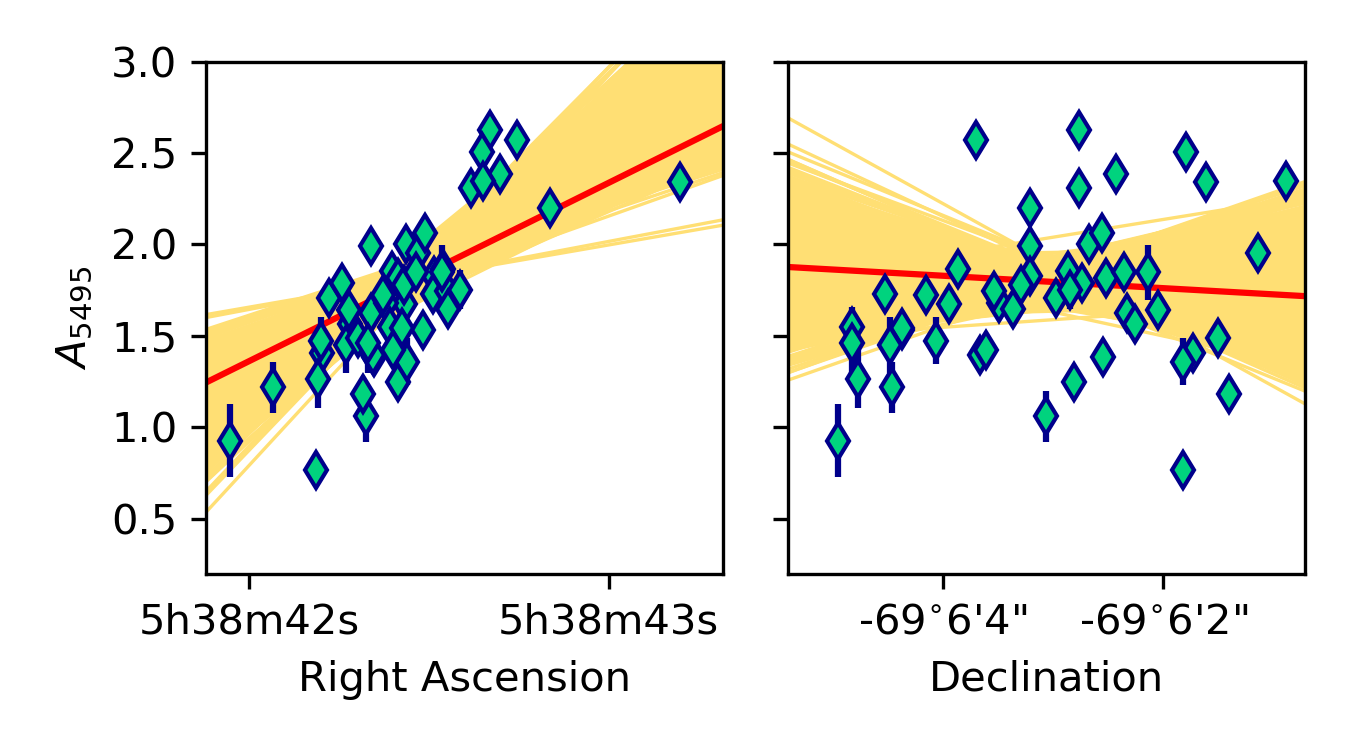}
    \caption{Extinction of the stars in the core of R136 ($A_{5495}$) plotted against their right ascension (left) and declination (right). Colours as in \cref{fig:c2_vs_Rv}. }
    \label{fig:spatial_trends_fit}    
\end{figure}

Upon plotting the slope in the optical and NIR, captured by $V-I$, versus the extinction $A_{5495}$, we see a strong correlation between the two (\cref{fig:c2_vs_Rv}):
\begin{equation}\label{eq:VI_AV}
A_{5495} = [0.38\pm0.23] + [3.86\pm0.58](V-I).
\end{equation}
Under the assumption that the dust properties of sightlines towards the outskirts of R136 are similar to those of sightlines towards the core, we can use this relation to estimate $A_{5495}$ for 1657 sources with $V~<~19$ in the catalogue of \citet{2011ApJ...739...27D} for which both $V$ and $I$ magnitudes are available. The magnitude cut-off ensures that we exclude the vast majority of the pre-main sequence stars included in the \citet{2011ApJ...739...27D} catalogue (see their Fig.~8). 
For the 1657 stars for which we estimate the extinction in this way, we find an average of $A_{5495} = 1.88 \pm 0.94$. The spatial distribution of the obtained extinction values is presented in \cref{fig:30dor_vfts_demarchi}, and shows that the trend in $A_{5495}$ that was observed throughout the cluster core continues east of the cluster core. We discuss the trends in extinction towards R136 relative to the larger field in the following subsection and in \cref{sec:dis:spatial}.

\subsection{Spatial trends in extinction \label{sec:res:spatial}}

We find a significant variation in extinction across the core of R136. What we refer to as the `core' spans a region of about 4" in diameter, corresponding to 1~pc for the LMC distance. Panel (a) of \cref{fig:2maps} shows a spatial map of extinction in the core. A gradient from east to west is clearly visible, where the east side of the cluster is more extincted than the west side. Inspecting the stars of the \citet{2011ApJ...739...27D} sample in \cref{fig:30dor_vfts_demarchi}, we see that this trend extends to outside the core. 

The extinction gradient is visualised quantitatively in \cref{fig:spatial_trends_fit}, where $A_{5495}$ values of the core stars are plotted as a function of right ascension (a higher value means more to the east) and declination (a lower value means more to the south). We see a particularly strong relation as a function of right ascension, but there seems to be no significant trend  as a function of declination. 
We can express the extinction $A_{5495}$ across the core of R136 as a function of spatial coordinates as follows:
\begin{equation}\label{eq:spatial}
    A_{5495} = [1.93\pm0.04] + [1.74\pm0.23](RA - 84.68)/0.006,     
\end{equation}
where $RA$ is the right ascension expressed in decimal degrees; as we do not find a significant trend as a function of declination, \cref{eq:spatial} only depends on right ascension.

Panel (b) of \cref{fig:2maps} shows another map of the core but this time with the values for \Rmono overplotted. We see a spatial trend for this quantity too,  where we find higher values of \Rmono towards the east of the cluster, although this trend is less clear than in the case of $A_{5495}$. 
Panel (c) of \cref{fig:2maps} shows that there is a strong correlation between \Rmono and $A_V$, and that the relation between the two implies $A_{4405}-A_{5495} \approx 0.4$. 
We note that if we refit the SED while adopting a fixed value of \Rmono (equal for all stars), we still recover the spatial trend for $A_{5495}$, although the gradient is slightly less steep. We discuss possible causes of the spatial trend in \cref{sec:dis:spatial}. 

\begin{figure*}
    \centering
    \includegraphics[width=1.0\textwidth]{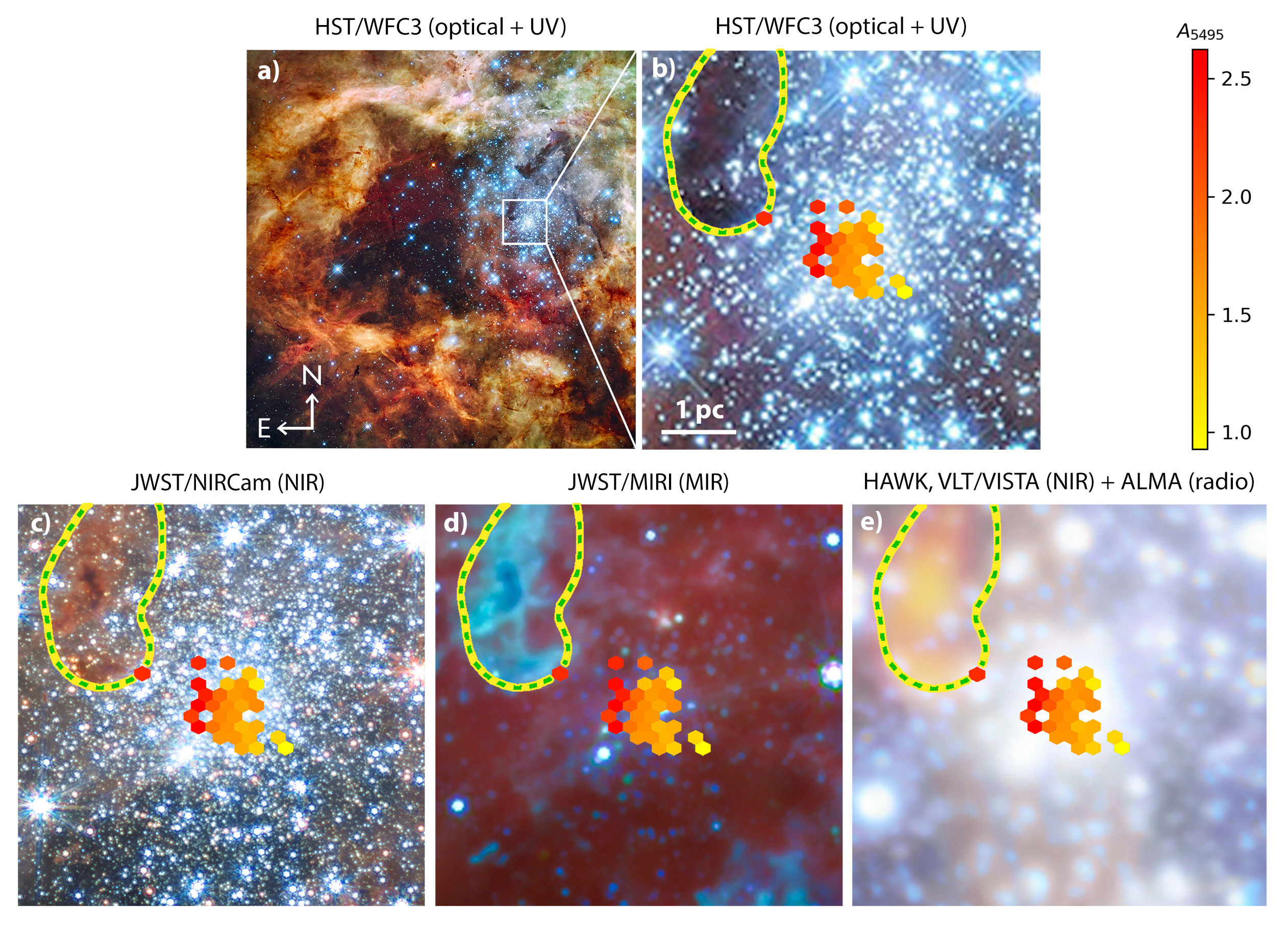}
    \caption{Multi-wavelength view of the cluster R136 and surroundings.  Panels \textit{a)} and \textit{b)} show the cluster imaged in optical and UV with WFC3/HST. The wider area around R316 is shown on the left, on the right we zoom in on the core of the cluster. The extinction map is projected onto the zoomed image in yellow to red colours (as in \cref{fig:2maps}). Imaged in optical and UV, the dark cloud north-east of the cluster, an extension of the Stapler Nebula, appears to have a considerable projected distance from the cluster core ($\approx1-2$~pc). The green-yellow dashed line indicates the contour of a $^{13}$CO column density of $10^{15}$~cm$^{-2}$ (see \cref{fig:13COmap}). 
    Panels \textit{c)}, \textit{d)} and \textit{e)} are as panel \textit{b)}, but with different background images: a NIR image captured with NIRCam on James Web Space Telescope (JWST), a mid-infrared (MIR) image  captured with MIRI on JWST, and a composite NIR/(sub-)mm image, captured in NIR with HAWK- I/VLT and VISTA, and in (sub-)mm wavelengths with ALMA. In panel \textit{c)} the orange-brown colours show cold gas corresponding roughly with the dark cloud in the HST image. Moving to longer wavelengths in panel \textit{c)}, the stars fade and the cool gas consisting of hydrocarbons is lighting up (turquoise). In panel \textit{e)} light pink regions correspond to relatively hot gas (NIR), and red-yellow areas indicate the presence of cold, dense gas (ALMA). 
    It seems possible that the extinction gradient in R136 may be associated with a lower density fringe of the extension of the Stapler Nebula; see also \cref{fig:13COmap} for additional evidence. 
    Image credits (from left to right): NASA, ESA, F. Paresce (INAF-IASF, Bologna, Italy), R. O'Connell (University of Virginia, Charlottesville), and the Wide Field Camera 3 Science Oversight Committee; NASA, ESA, CSA, and STScI; IMAGE: NASA, ESA, CSA, STScI, Webb ERO Production Team; ESO, ALMA (ESO/NAOJ/NRAO)/Wong et al., ESO/M.-R. Cioni/VISTA Magellanic Cloud survey, Acknowledgment: Cambridge Astronomical Survey Unit.}
    \label{fig:HSTimge_pretty}
\end{figure*}

\begin{figure}
    \centering
    \includegraphics[width=0.43\textwidth]{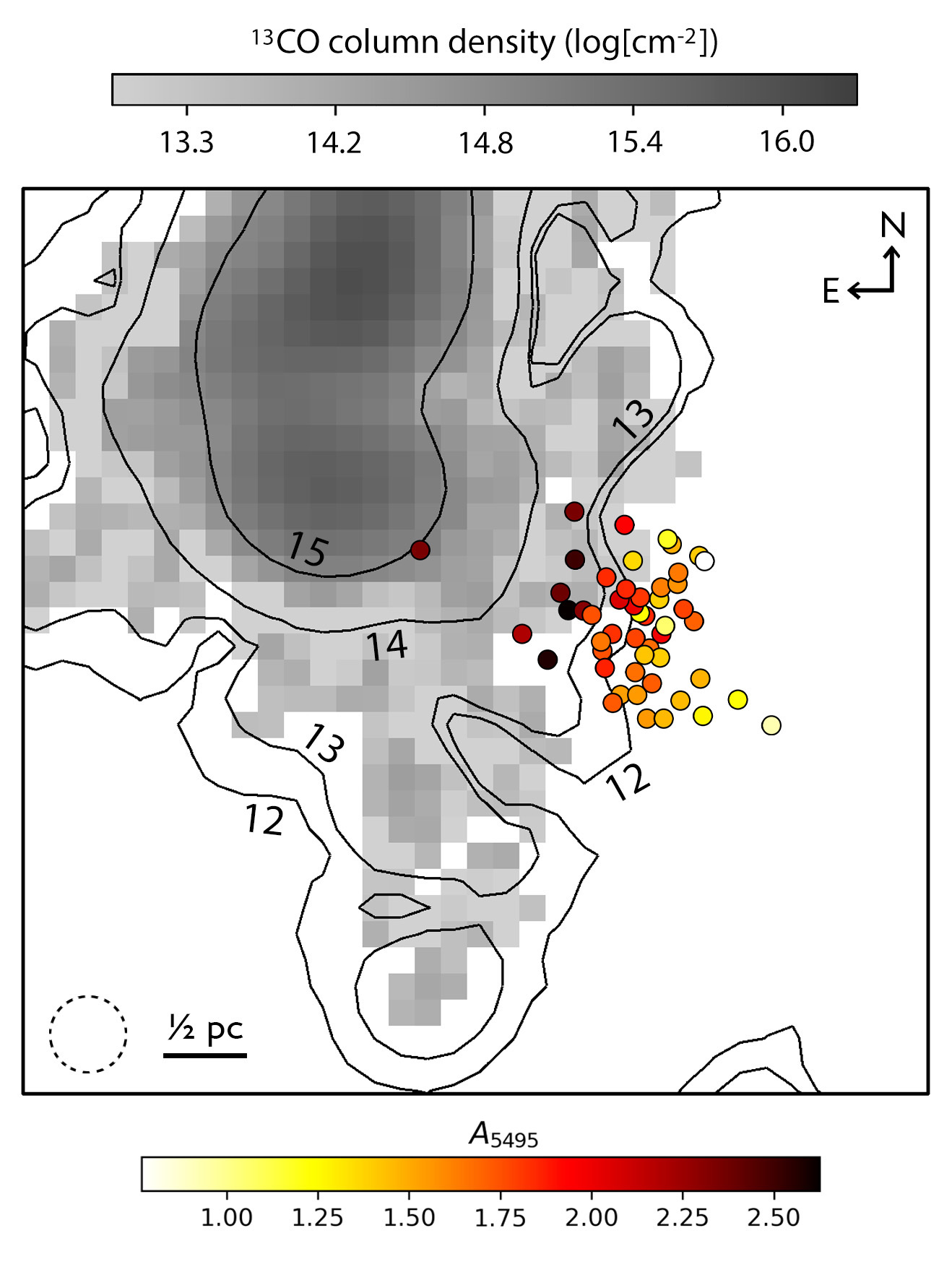}
    \caption{$^{13}$CO column density map of R136 and surroundings, obtained with ALMA \citep[][]{2022ApJ...932...47W}. Darker regions indicate a higher $^{13}$CO column density; in white regions no $^{13}$CO was detected. Contours, created with DS9 (smoothing = 4), correspond to lines of constant $^{13}$CO column density, the value of which is indicated in the plot in units of $\log[\mathrm{cm}^{-2}]$. The dashed circle in the lower left corner of the image indicates the beam size. Small filled circles indicate the positions of the stars in the R136 core; the colour of each circle corresponds to the measured extinction, with darker colours corresponding to a higher extinction.  Comparing the extinction of the R136 core stars with the $^{13}$CO map, we see that in regions where $^{13}$CO is detected we measure stronger extinction.  \label{fig:13COmap}}
\end{figure}

\subsection{The UV curvature ($c_4$) \label{sec:uv_curvature}}

For the parameter $c_4$, describing the curvature of the extinction curve in the far-UV ($x > 5.9~\mu\mathrm{m}^{-1}$ or $\lambda < 1695~ \mathrm{\AA}$), we find an average value of $c_4 = 0.09\pm0.08$. 
This value is lower than found by \citet{1999ApJ...515..128M} and \citet{2003ApJ...594..279G} for the supergiant shell LMC SGS~2 near 30 Doradus; these authors find $c_4 = 0.42\pm0.08$ and $c_4 = 0.29\pm0.06$, respectively. 
A low value of $c_4$ corresponds to a weak far-UV curvature. 

We note that the value of $c_4$ we find for the R136 core stars is possibly even lower than the value quoted above. This is due to possible (relative) flux-calibration issues of the STIS spectra, as revealed in \cref{fig:integrated}: upon comparing the integrated flux of the R136 core as recorded by STIS with the flux as recorded by GHRS, the two start diverging at around $\lambda < 1400$~\AA. In this wavelength regime, which is especially sensitive to the value of $c_4$, the STIS flux is lower than the GHRS flux, namely by a factor of  about 0.88. If the low fluxes are due to an unknown calibration issue of the STIS spectra, and in reality the shape of the spectra is more like that recorded by GHRS, then in our analysis we have overestimated $c_4$. 
In order to match the reddened model fluxes to the GHRS flux (i.e., for the reddened model fluxes to be a factor $1/0.88=1.14$ higher than the STIS spectra, at $\lambda = 1280$~\AA), the value of $c_4$ would have to approach zero, that is, no far-UV curvature at all.

\section{Discussion\label{sec:discussion}}

\subsection{Extinction due to an extension of the Stapler Nebula\label{sec:dis:spatial}}

We find a strong gradient in the amount of extinction across the 1~pc core of the R136 cluster, ranging from $A_{5495} \approx 1.0$ in the west, up to $A_{5495} \approx 2.6$ east of the cluster core. The estimated extinction of the stars in the outskirts of R136 suggests that this trend extends outside the cluster core (\cref{fig:30dor_vfts_demarchi}). 
We compare the extinction gradient with images of the cluster in order to identify the larger-scale structure of the intervening gas and dust. 
In the composite optical and UV HST image of the cluster and surroundings (\cref{fig:HSTimge_pretty}, top row), we see a dark cloud to the northeast of the cluster. The cloud is an extension of the Stapler Nebula; \citet{2018ApJ...852...71K} study the nebula and its relation to R136 in detail. The fact that there are hardly any stars visible in front of this cloud suggests that it is situated in the foreground \citep{2018ApJ...852...71K,2022ApJ...932...47W}. This cloud is even more clearly visible in the IR images captured by NIRCam and MIRI on the James Webb Space Telescope (JWST, \cref{fig:HSTimge_pretty}, bottom row). These high-resolution images reveal the detailed structure of the cloud, with one arm of the cloud extending all the way to the easternmost star of our sample, for which we find a relatively high extinction. 
Nonetheless, seen in the NIR, optical, and UV, the west edge of the cloud has a projected distance from the centre of R136 of $\approx1-2$~pc, and these images therefore do not provide direct evidence that the cloud is responsible for the extinction gradient across the cluster. 

However, images at even longer wavelengths reveal that the molecular cloud stretches out farther than can be seen in the HST and JWST images. \citet[][]{2022ApJ...932...47W} observed 30 Doradus with ALMA, and map the strength of the CO(2 - 1) rotational line, a tracer of cold molecular gas. With high spatial resolution and sensitivity, ALMA (beam size: 1"75) allows us to distinguish details of the distribution of the cold gas. The cluster core spans about 2.3$\times$ the beam size, allowing us to distinguish between the  $^{13}$CO column density in the east versus the west side of the cluster. 
\cref{fig:13COmap} shows a map of the column density of $^{13}$CO, as well as the positions of the R136 core stars, marked by their extinction. It is clear that in regions where $^{13}$CO is detected (eastern half of the cluster core), the extinction is higher, suggesting that this molecular cloud contains the dust grains that are causing the extinction gradient. 
Furthermore, we see that the position of the highest $^{13}$CO column density ($\geq10^{15}$~cm$^{-2}$) corresponds to the position of the dark cloud in \cref{fig:HSTimge_pretty} (yellow-green dashed line). The molecular gas mapped by the CO(2 - 1) line therefore seems to be associated with the dark cloud of \cref{fig:HSTimge_pretty}.

Having addressed the spatial trend in $A_V$, we now turn to the trend we find for \Rmono. On average, we find higher values of \Rmono towards the east of the cluster than towards the west side. Higher \Rmono values are interpreted as the result of fewer very small grains (having sizes $<250~\AA$) of either silicate particles \citep{2017ApJ...835..107X} or a mixture of graphitic and silicate grains \citep{2001ApJ...548..296W}, relative to larger grains. As \Rmono has a strong positive correlation with $A_V$ (see panel \textit{c)} of \cref{fig:2maps}), we can see from \cref{fig:13COmap} that higher values of \Rmono map to higher values of $^{13}$CO column density. This could indicate that grain growth through aggregation is more efficient  in the denser
environments of the molecular cloud, or alternatively that denser cloud regions shield the grains more efficiently from strong radiation fields that may break apart dust particles. 

Very limited information is available on the relation between $A_{5495}$ and \Rmono in other star forming regions in the LMC. \citet{2018A&A...613A...9M} study the extinction towards O-type stars in Galactic H{\sc ii} regions, and find results that are not in line with ours. These authors also find differences in \Rmono on relatively small spatial scales ($\approx 1-8$~pc, $A_{5495} = 2.0-7.3$), but higher values of \Rmono do not always coincide with higher values of $A_{5495}$. 
On the contrary, combining the measurements of four different H{\sc ii} regions (shown in their Figure~7), a tentative opposite trend appears, with larger values of \Rmono generally corresponding to lower $A_{5495}$.

\subsection{Revised WNh star masses \label{sec:WNhmass}}

\renewcommand{\arraystretch}{1.3}
\begin{table}[]
    \centering
    \caption{Revised ages and masses of the WNh stars with 1$\sigma$ errors based on the stellar parameters derived by \citet{2022arXiv220211080B} and accounting for the extinction values obtained in this work. \label{tab:newmassesWNH}}
    \begin{tabular}{l l l l }
    \hline \hline 
                & $\phantom{-}$R136a1 & $\phantom{-}$R136a2 & $\phantom{-}$R136a3 \\ 
                 \hline 
${M}_{\rm act}$ (\Msun) & $237 \pm^{28}_{22}$  & $140 \pm^{10}_{9}$  & $174 \pm^{18}_{10}$ \\
${M}_{\rm ini}$ (\Msun) & $277 \pm^{24}_{27}$  & $171 \pm^{11}_{7}$  & $209 \pm^{10}_{9}$ \\
Age (Myr) & $0.94 \pm^{0.17}_{0.16}$  & $1.64 \pm^{0.13}_{0.12}$  & $1.32 \pm^{0.19}_{0.25}$ \\
log $L/L_\odot$ & $6.82 \pm^{0.06}_{0.04}$  & $6.59 \pm^{0.02}_{0.03}$  & $6.69 \pm^{0.02}_{0.02}$ \\
\hline
    \end{tabular}
\end{table}

\begin{figure*}
    \centering
    \includegraphics[width=0.9\textwidth]{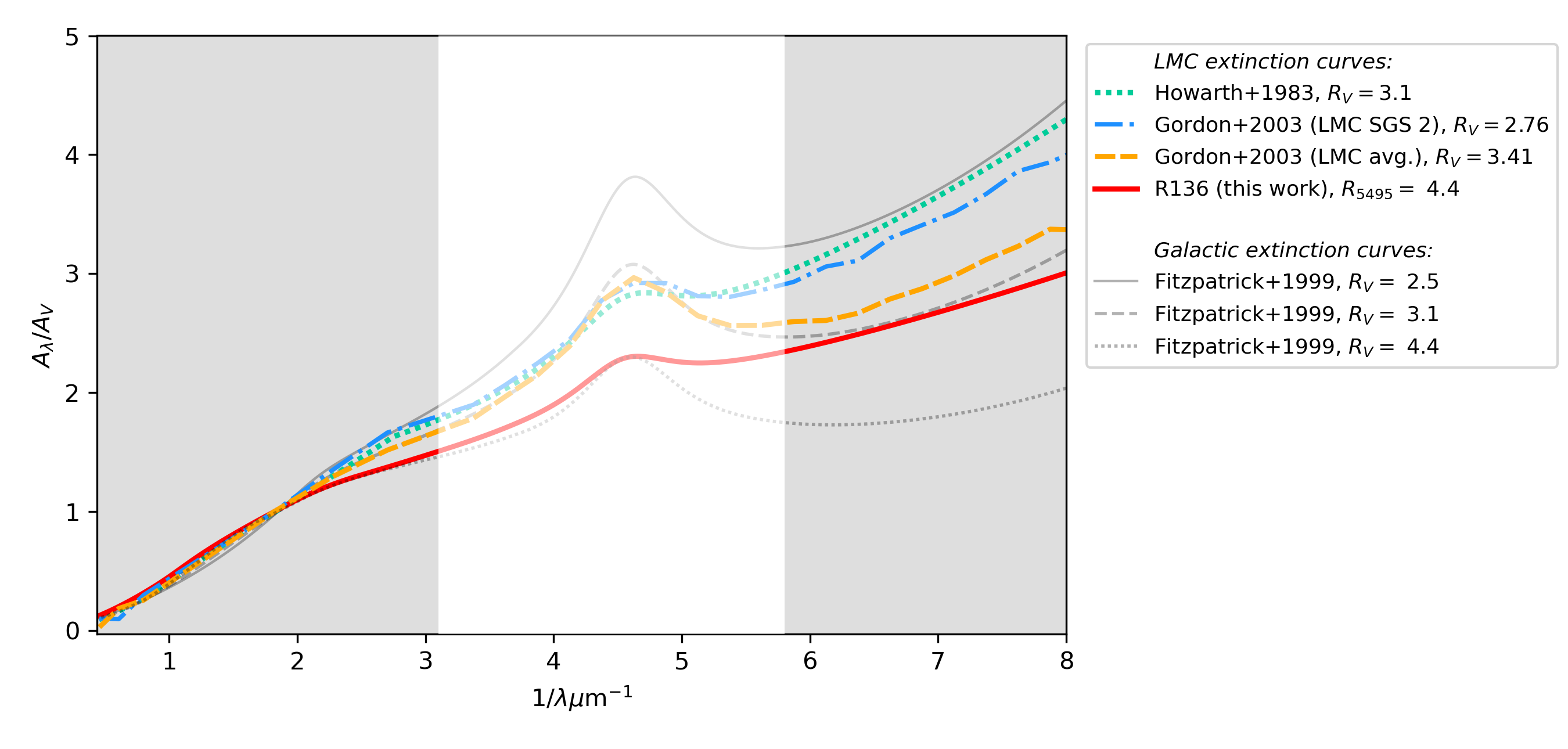}
    \caption{Comparison of extinction curves tailored to (specific regions within) the LMC. Shown are the average extinction curve of \citet[][green dotted line]{1983MNRAS.203..301H}, which is nearly equal to that of \citet[][not shown]{1992ApJ...395..130P} and two curves of \citet{2003ApJ...594..279G}, based on samples within LMC SGS~2 (blue dash-dotted line), and a sample in other parts of the LMC (`LMC avg.', orange dashed line). We compare these curves to the average curve derived in this work, which we show for $R_{5495} = 4.4$ (the average value towards R136; red solid line). 
    For reference, we also show the Galactic curve of \citet{1999PASP..111...63F} for different values of $R_V$ (grey lines).
    The grey areas indicate parts of the SED that we analysed in this work. The white area contains the \bump, a wavelength region that was not covered by our data; the extinction curves are also shown in this wavelength range for completeness, but we highlight the fact that the bump parameters of our R136 curve are simply taken from the LMC SGS~2 curve of \citet{2003ApJ...594..279G}.
    \label{fig:comp_all_curves}}
    
\end{figure*}

\begin{figure}
    \centering
    \includegraphics[width=0.49\textwidth]{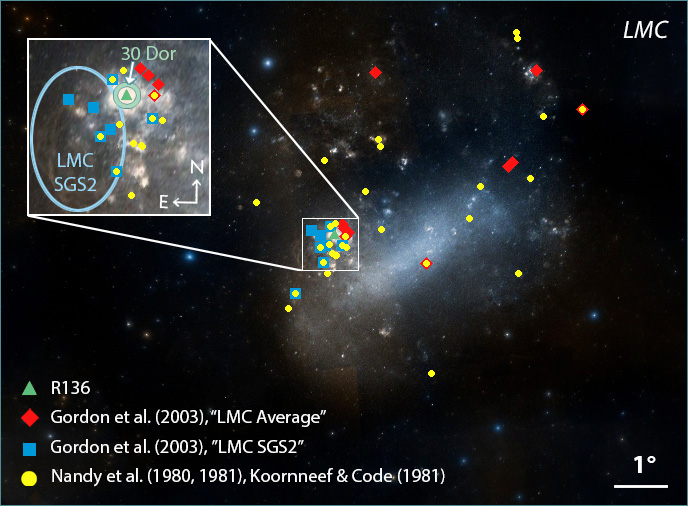}
    \caption{Position of stars of different samples that were used to derive extinction curves towards the LMC. The samples of \citet{1980Natur.283..725N,1981MNRAS.196..955N} and \citet[][yellow circles]{1981ApJ...247..860K} were used to construct the curves of \citet{1983MNRAS.203..301H} and \citet{1992ApJ...395..130P}. \citet{2003ApJ...594..279G} split their sample into two and derive a separate curve for each, referring to one sample as LMC2 (LMC SGS~2, blue squares) and the other as `LMC Average' (red diamonds). The green triangle denotes the position of R136. This figure was made with use of the Aladin Sky Atlas \citep{2000A&AS..143...33B}; the background consists of DSS2 colour images. }
    \label{fig:lmc_samples_map}
\end{figure}

When accounting for the new values of the extinction, we find changes in luminosity of the R136 core stars of on average $0.03$~dex compared to the luminosities obtained by \citet{2022arXiv220211080B}; the revisions in $\log L/L_\odot$ for individual stars range from $-0.13$ to $+0.04$~dex. 
The changes in luminosity imply small changes of the stellar masses. For the O-stars in the core of R136, we estimate the mass-reduction by eye using their positions on the Hertzsprung–Russell diagram (HRD) and the tracks of \citet{2011A&A...530A.115B} and \citet{2015A&A...573A..71K}\footnote{The HRD for the core of R136 with tracks of \citet{2011A&A...530A.115B} and \citet{2015A&A...573A..71K} can be found in Fig. 10 of \citet{2022arXiv220211080B}.}. We estimate that the stellar mass is typically changed by $5\%-10\%$. 

We performed a detailed analysis of the masses of the most massive stars in the sample, namely R136a1, R136a2, and R136a3. We do this in the same manner as in \citet{2022arXiv220211080B}, only now with our new luminosity values. 
For this, we use the Bayesian tool {\sc Bonnsai}\footnote{The Bonnsai web-service is available at \url{https://www.astro.uni-bonn.de/stars/bonnsai/}.} \citep{2017A&A...598A..60S}, in combination with the evolutionary models of \citet{2011A&A...530A.115B} and \citet{2015A&A...573A..71K}. {\sc Bonnsai} allows the comparison of observed stellar parameters with stellar evolution models in order to infer posterior distributions of model parameters such as initial and current stellar mass. For our derivations of stellar masses and ages we use temperature, helium abundance, and surface gravity as derived by \citet{2022arXiv220211080B}, and the revised luminosity from this work. As priors, we adopt the \citet{1955ApJ...121..161S} initial mass function and the rotation distribution of \citet{2013A&A...560A..29R}. 

We find that the evolutionary and initial masses we derive for R136a1 and R136a3 agree within uncertainties with those of \citet{2022arXiv220211080B}, while for R136a2 our masses are 25\% lower. The new age that we derive for R136a3 agrees well with that derived by \citet{2022arXiv220211080B}, and for R136a1 and R136a2 we find downward and upward revisions for the age of about 20\%, respectively. We note that \citet{2022arXiv220211080B} provide two sets of values for the stellar parameters; we adopt the values of their optical~+~UV fits. 
The new values can be found in \cref{tab:newmassesWNH}. 
Comparing our initial masses with literature values, we find that they are generally higher than those derived by \citet{2017IAUS..329..131R} and \citet{2022ApJ...935..162K}, although they do agree with the latter  within errors. 
We note that initial masses of these stars, both ours and those presented in literature, are highly model dependent (see, e.g. \citealt{2021A&A...647A..13G}, \citealt{2022MNRAS.516.4052H}, and \citealt{2022arXiv220211080B}, their Fig. 17). Moreover, even within a given set of evolution models, the derived initial mass is sensitive to the combination of observables used for the comparison with the evolutionary models. Lastly, all these analyses are based on single-star evolution scenarios, but it is important to keep in mind that these very massive stars might be merger products \citep[e.g.][]{2012MNRAS.426.1416B}. 

\subsection{Anomalous extinction towards R136? \label{sec:dis:RV}}

 The average $R_V$ value we find for R136 ($R_V \approx R_{5495} = 4.38\pm 0.87$) is in good agreement with previously obtained averages for the region\footnote{In this section, we use the broadband and the monochromatic definitions of the total-to-relative extinction, \RV and \Rmono, interchangeably: as the sources that we discuss are not heavily reddened, and their SEDs are fairly similar, the differences between them are assumed to be sufficiently small.} \citep[][for R136, and \citealt{2013AA...558A.134D,2014AA...570A..38B,2014AA...564A..63M,2014MNRAS.445...93D,2016MNRAS.455.4373D}, for the wider 30 Doradus region]{2016MNRAS.458..624C,2020MNRAS.499.1918B}. Such high values of $R_V$ are, in the Milky Way, associated with low UV extinction, but this is not what we observe for R136. 
The UV extinction for $R_V = 4.4$ in R136 is far higher than the UV extinction in the Galactic case for the same $R_V$ (\cref{fig:comp_all_curves}). 
From a physical point of view, it is not unexpected that the slope in the optical ($R_V$) and the strength of UV extinction can vary independently, as these two parts of the curve are thought to be associated with complementary dust particle populations \citep[e.g.][]{2001ApJ...548..296W,2017ApJ...835..107X}. Nevertheless, one population can be related to the other, as in the Milky Way their relative contributions do not vary randomly but follow a relatively simple relation (i.e. \cref{eq:fitz_uv_orig}). This relation is dictated by the properties of the interstellar dust, which in turn are affected by the environment. In 30 Doradus, or at least in R136, we do not observe this relation between \RV and the UV extinction. 
Possibly, the dust in and near 30 Doradus is affected by the intense radiation and mechanical energy that is being deposited in the interstellar medium by the large population of hot stars in the region, as well as by previous powerful supernova explosions of local massive stars \citep{2019ApJ...878...31D}. 

In \cref{fig:comp_all_curves}, we compare our $R_V$-dependent extinction law towards R136 with other extinction curves that are tailored to regions of the LMC. Contrasting the average curve towards R136 with the curves of \citet{2003ApJ...594..279G}, it appears that, similar to Galactic sightlines, higher values of \RV correspond to lower values of UV extinction. This would suggest that a relation between \RV and UV extinction may exist, but that it differs from the Galactic relation. The fact that we do not observe this relation within R136 could be related to the uncertainties in the UV flux calibration, which increase the scatter on our measured $c_2$ values. On the other hand, \citet[][]{1983MNRAS.203..301H}, who also derive an LMC average curve and obtain $R_V = 3.1$  for their sample, find a UV extinction that is too strong with respect to
their $R_V$  to match the tentative relation between \RV and UV extinction that the other curves in \cref{fig:comp_all_curves} might suggest. Furthermore, the result of \citet[][]{1983MNRAS.203..301H} appears to be in contradiction with the results of \citet{2003ApJ...594..279G}. 

The differences between the `LMC average' curves of \citet{2003ApJ...594..279G} and \citet{1983MNRAS.203..301H} can likely be understood by the fact that their samples consist of different sightlines. \citet[][and also \citealt{1992ApJ...395..130P}, who find a very similar average curve,]{1983MNRAS.203..301H} use the samples of \citet{1980Natur.283..725N,1981MNRAS.196..955N}, and \citet[][]{1981ApJ...247..860K}, of which about half of the sightlines lie around 30 Doradus, with the other half being spread out over other regions in the LMC. For their LMC average curve,  \citet{2003ApJ...594..279G}
use  sightlines that are more clustered and lower in number. The sightlines of the different LMC samples are shown in \cref{fig:lmc_samples_map}.  The sample used to derive the curve towards LMC SGS~2 is also indicated, as are R136 and 30 Doradus. We note that while sightlines near 30 Doradus are considered in all samples, no sightline towards 30 Doradus itself is included in any of the samples; the present work on R136 is therefore unique in this respect. 

In any case, we cannot draw firm conclusions on the $R_V$ dependence of UV extinction with only four curves. It therefore remains unclear as to whether or not there exists a relation between the different grain populations in the LMC, or what the nature of this relation  could be. It would be of value to carry out an LMC-wide study of the $R_V$ dependence of UV extinction in order to investigate this further.  

\renewcommand{\arraystretch}{1.}

\section{Conclusion\label{sec:conclusion}}

Employing the extinction-without-standards method, we infer NIR to UV extinction characteristics towards 50 stars in the core of R136. 
On average, we find an extinction of $A_V \approx A_{5495} = 1.70\pm0.45$. However, we infer a strong spatial gradient in extinction properties across the cluster core, where the extinction in the east is about one magnitude higher than the extinction in the west of the cluster. Comparing our extinction map to multi-wavelength observations of the same region, we conclude that the observed extinction gradient is likely caused by material belonging to an extension of a molecular cloud called the Stapler Nebula, which lies to the northeast of the cluster and stretches all the way to the R136 core. 

In line with previous studies, we obtain a relatively high average value of $R_V \approx R_{5495} = 4.38\pm 0.87$ towards R136. Moreover, we find that the UV extinction towards R136 is significantly stronger than the canonical Galactic extinction at the same value for the same \RV, implying a relatively large fraction of small particles near R136. The intense radiation field and mechanical energy that is being deposited in the 30 Doradus interstellar medium by the hot stars and their powerful core-collapse supernovae could play a role in this process. 
A consequence of the stronger UV absorption is that less ionising photons can escape. 
At $A_V = 1.0$, the extinction towards R136 at UV wavelengths in the range $\lambda \approx 1700 - 1250$~\AA\xspace is about one magnitude higher compared to the canonical Galactic extinction at the same \RV and $A_V$, implying that the fraction of ionising photons that can escape is a factor 2.5 lower ($e^{\tau} \approx e^{A_V} \approx 0.4$). 

We have now investigated the relation between $R_V$ and extinction in the UV for one region in the LMC, namely R136. Extending this investigation to different environments throughout the Magellanic Clouds would be an interesting topic for future studies: 
knowledge about the interdependence of different dust populations 
could provide insights into the environmental factors determining the properties and evolution of the dust therein. This would be of particular interest in the context of starburst galaxies, where stellar populations are unresolved and dust properties need to be known in order to interpret observations. 
This includes the role of dust in star-bursting regions in absorbing and reprocessing of ionising photons trying to escape to inter-galactic space.  

\begin{acknowledgements}

We thank the referee Jes\'{u}s Ma\'{i}z Apell\'{a}niz for providing constructive comments. This publication is part of the project `Massive stars in low-metallicity environments: the progenitors of massive black holes' with project number OND1362707 of the research TOP-programme, which is (partly) financed by the Dutch Research Council (NWO). Observations were taken with the NASA/ESA \textit{HST}, obtained from the data archive at the Space Telescope Institute. This research has made use of the SIMBAD database, operated at CDS, Strasbourg, France \citep{2000A&AS..143....9W}. 

\end{acknowledgements}

\bibliographystyle{aa}
\bibliography{references}

\begin{appendix}

\section{UV bands\label{tab:appUVbins}}

\Cref{tab:app:wavesUV} lists the wavelength ranges that we used for binning our UV flux measurements. We use the same ranges for all stars. The average flux of each wavelength interval is the value that was used for the fitting. This is equivalent to constructing a (synthetic) passband with a transmission of 1.0 in the wavelength range corresponding to the filter, and 0.0 outside that range. 

\begin{table}[h!]
    \centering
    \caption{Minimum and maximum wavelengths ($\lambda_{\rm min}$ and $\lambda_{\rm max}$) of synthetic passbands used for the UV, named by their (rounded) central wavelength.}
    \label{tab:app:wavesUV}
\begin{tabular}{l l l l l l l l l}
\hline\hline
Passband & $\lambda_{\rm min}$ ({\rm \AA}) & $\lambda_{\rm max}$ ({\rm \AA}) \\ 
\hline 
$\lambda$1280 & 1265.00 & 1295.00\\
$\lambda$1319 & 1307.00 & 1330.00\\
$\lambda$1351 & 1347.00 & 1355.00\\
$\lambda$1420 & 1405.00 & 1435.00\\
$\lambda$1455 & 1440.00 & 1470.00\\
$\lambda$1498 & 1475.00 & 1520.00\\
$\lambda$1585 & 1570.00 & 1600.00\\
$\lambda$1618 & 1612.00 & 1624.00\\
$\lambda$1688 & 1675.00 & 1700.00\\
\hline 
\end{tabular}
\end{table}

\section{Best-fit values for individual sources}

The best-fit values for individual sources in the core of R136 can be found in \cref{tab:app:R136results}. 

\renewcommand{\arraystretch}{1.11}    
\begin{table*}[]
\caption{Best-fit values of the R136 core stars$\dagger$.}
\begin{tabular}{l r@{$\pm$}l r@{$\pm$}l l l r@{$\pm$}l r@{$\pm$}l l}
\hline \hline
Source  & \multicolumn{2}{l}{$R_V$} & \multicolumn{2}{l}{$A_V$} & $A_K$ & $\log L/L_\odot$ & \multicolumn{2}{l}{$c_2$} & \multicolumn{2}{l}{$c_4$}  & $\chi^2_\mathrm{red}$\\ \hline
R136a1 & 4.25 & 0.10 & 1.85 & 0.02 & 0.24 & $6.82\pm0.01$ & 1.28 & 0.05 & 0.00 & $^{0.12}_{0.00}$ & 4.44 \\
R136a2 & 2.93 & 0.15 & 1.25 & 0.04 & 0.14 & $6.59\pm0.01$ & 1.62 & 0.11 & 0.00 & $^{0.34}_{0.00}$ & 14.32 \\
R136a3 & 4.63 & 0.08 & 1.99 & 0.02 & 0.26 & $6.69\pm0.01$ & 1.14 & 0.03 & 0.07 & $^{0.12}_{0.07}$ & 1.81 \\
R136a4 & 3.30 & 0.14 & 1.38 & 0.03 & 0.17 & $6.23\pm0.01$ & 1.43 & 0.07 & 0.00 & $^{0.42}_{0.00}$ & 4.83 \\
R136a5 & 4.67 & 0.31 & 2.00 & 0.05 & 0.27 & $6.34\pm0.01$ & 1.33 & 0.13 & 0.00 & $^{0.02}_{0.00}$ & 13.18 \\
R136a7 & 5.04 & 0.61 & 1.82 & 0.07 & 0.25 & $6.29\pm0.02$ & 1.87 & 0.30 & 0.00 & $^{0.69}_{0.00}$ & 24.60 \\
R136b & 5.66 & 0.21 & 2.57 & 0.04 & 0.36 & $6.39\pm0.01$ & 1.18 & 0.08 & 0.06 & $^{0.20}_{0.06}$ & 8.72 \\
H30 & 4.13 & 0.13 & 1.63 & 0.03 & 0.21 & $5.72\pm0.01$ & 1.15 & 0.06 & 0.06 & $^{0.15}_{0.06}$ & 3.90 \\
H31 & 3.94 & 0.24 & 1.53 & 0.05 & 0.20 & $5.93\pm0.02$ & 1.24 & 0.13 & 0.06 & $^{0.39}_{0.06}$ & 19.94 \\
H35 & 4.66 & 0.29 & 1.57 & 0.04 & 0.21 & $5.81\pm0.01$ & 1.34 & 0.09 & 0.05 & $^{0.20}_{0.05}$ & 5.69 \\
H36 & 5.79 & 0.31 & 2.63 & 0.05 & 0.37 & $6.31\pm0.02$ & 1.15 & 0.12 & 0.03 & $^{0.35}_{0.03}$ & 17.16 \\
H40 & 5.55 & 0.41 & 1.95 & 0.06 & 0.27 & $5.93\pm0.02$ & 1.30 & 0.15 & 0.11 & $^{0.41}_{0.11}$ & 20.02 \\
H45 & 5.38 & 0.13 & 2.20 & 0.02 & 0.30 & $5.82\pm0.01$ & 1.25 & 0.06 & 0.13 & $^{0.19}_{0.13}$ & 3.39 \\
H46 & 5.04 & 0.32 & 2.38 & 0.06 & 0.32 & $6.12\pm0.02$ & 1.18 & 0.19 & 0.13 & $^{0.64}_{0.13}$ & 23.42 \\
H47 & 5.47 & 0.52 & 2.50 & 0.09 & 0.35 & $5.99\pm0.03$ & 1.15 & 0.31 & 0.22 & $^{1.01}_{0.22}$ & 39.86 \\
H48 & 5.42 & 0.37 & 2.31 & 0.06 & 0.32 & $5.98\pm0.01$ & 1.21 & 0.13 & 0.06 & $^{0.35}_{0.06}$ & 13.09 \\
H50 & 4.17 & 0.10 & 1.83 & 0.02 & 0.24 & $5.84\pm0.00$ & 1.11 & 0.04 & 0.09 & $^{0.11}_{0.09}$ & 1.59 \\
H52 & 4.80 & 0.23 & 1.67 & 0.03 & 0.23 & $5.69\pm0.01$ & 1.29 & 0.10 & 0.14 & $^{0.24}_{0.14}$ & 4.87 \\
H55 & 4.66 & 0.38 & 1.73 & 0.06 & 0.23 & $5.76\pm0.02$ & 1.27 & 0.16 & 0.07 & $^{0.44}_{0.07}$ & 18.26 \\
H58 & 4.96 & 0.41 & 2.06 & 0.06 & 0.28 & $5.87\pm0.01$ & 1.31 & 0.18 & 0.15 & $^{0.48}_{0.15}$ & 13.36 \\
H62 & 4.54 & 0.17 & 1.68 & 0.02 & 0.22 & $5.61\pm0.01$ & 1.18 & 0.06 & 0.21 & 0.14 & 1.86 \\
H64 & 5.23 & 0.45 & 2.35 & 0.08 & 0.32 & $5.85\pm0.02$ & 1.29 & 0.21 & 0.04 & $^{0.60}_{0.04}$ & 35.26 \\
H66 & 4.06 & 0.17 & 1.78 & 0.03 & 0.23 & $5.65\pm0.01$ & 1.12 & 0.07 & 0.25 & 0.19 & 2.27 \\
H68 & 5.21 & 0.13 & 2.34 & 0.02 & 0.32 & $5.69\pm0.01$ & 1.20 & 0.07 & 0.03 & $^{0.22}_{0.03}$ & 4.19 \\
H69 & 4.92 & 0.59 & 1.75 & 0.07 & 0.24 & $5.45\pm0.01$ & 1.46 & 0.14 & 0.08 & $^{0.14}_{0.08}$ & 1.19 \\
H70 & 4.18 & 0.45 & 1.85 & 0.08 & 0.24 & $5.68\pm0.02$ & 1.34 & 0.28 & 0.19 & $^{0.88}_{0.19}$ & 17.69 \\
H71 & 4.45 & 0.82 & 1.55 & 0.11 & 0.20 & $5.44\pm0.03$ & 1.35 & 0.33 & 0.07 & $^{0.76}_{0.07}$ & 43.68 \\
H73 & 3.68 & 0.23 & 1.65 & 0.04 & 0.21 & $5.15\pm0.01$ & 0.96 & 0.07 & 0.16 & 0.16 & 3.09 \\
H75 & 3.72 & 0.12 & 1.41 & 0.02 & 0.18 & $5.45\pm0.00$ & 1.40 & 0.06 & 0.17 & 0.14 & 1.34 \\
H78 & 4.59 & 0.24 & 1.71 & 0.03 & 0.23 & $5.50\pm0.01$ & 1.45 & 0.09 & 0.07 & $^{0.20}_{0.07}$ & 2.17 \\
H80 & 3.91 & 0.21 & 1.40 & 0.03 & 0.18 & $5.09\pm0.01$ & 1.19 & 0.08 & 0.00 & $^{0.19}_{0.00}$ & 2.30 \\
H86 & 2.36 & 0.60 & 1.06 & 0.14 & 0.11 & $5.42\pm0.03$ & 1.22 & 0.35 & 0.00 & $^{0.02}_{0.00}$ & 56.89 \\
H90 & 4.85 & 0.15 & 1.79 & 0.02 & 0.24 & $5.34\pm0.00$ & 1.31 & 0.05 & 0.03 & $^{0.09}_{0.03}$ & 0.52 \\
H92 & 4.66 & 0.16 & 1.64 & 0.02 & 0.22 & $5.21\pm0.00$ & 1.27 & 0.05 & 0.01 & $^{0.10}_{0.01}$ & 0.63 \\
H94 & 5.15 & 0.26 & 1.87 & 0.03 & 0.26 & $5.39\pm0.01$ & 1.39 & 0.08 & 0.04 & $^{0.17}_{0.04}$ & 1.44 \\
H108 & 1.91 & 0.26 & 0.77 & 0.06 & 0.07 & $4.88\pm0.01$ & 1.21 & 0.15 & 0.08 & $^{0.33}_{0.08}$ & 9.05 \\
H112 & 4.19 & 0.98 & 1.85 & 0.15 & 0.24 & $5.08\pm0.03$ & 1.16 & 0.35 & 0.10 & $^{0.72}_{0.10}$ & 24.59 \\
H114 & 4.64 & 0.69 & 1.55 & 0.08 & 0.21 & $5.21\pm0.02$ & 1.40 & 0.24 & 0.00 & $^{0.70}_{0.00}$ & 11.09 \\
H116 & 4.85 & 1.57 & 1.45 & 0.15 & 0.20 & $4.86\pm0.03$ & 1.54 & 0.54 & 0.07 & $^{1.01}_{0.07}$ & 37.99 \\
H120 & 2.36 & 0.88 & 0.93 & 0.20 & 0.09 & $4.78\pm0.06$ & 1.24 & 0.57 & 0.00 & $^{1.19}_{0.00}$ & 164.30 \\
H121 & 4.33 & 1.39 & 1.46 & 0.17 & 0.19 & $4.77\pm0.03$ & 1.24 & 0.48 & 0.23 & $^{1.04}_{0.23}$ & 48.33 \\
H123 & 3.39 & 0.87 & 1.22 & 0.14 & 0.15 & $4.87\pm0.03$ & 1.37 & 0.44 & 0.03 & $^{1.06}_{0.03}$ & 59.44 \\
H132 & 3.79 & 0.51 & 1.49 & 0.08 & 0.19 & $5.01\pm0.01$ & 1.22 & 0.19 & 0.10 & $^{0.37}_{0.10}$ & 7.05 \\
H134 & 4.16 & 1.38 & 1.27 & 0.16 & 0.16 & $4.72\pm0.04$ & 1.46 & 0.62 & 0.09 & $^{1.43}_{0.09}$ & 62.54 \\
H135 & 2.98 & 0.69 & 1.36 & 0.13 & 0.16 & $4.65\pm0.02$ & 0.78 & 0.20 & 0.32 & $^{0.33}_{0.32}$ & 10.99 \\
H139 & 4.67 & 1.13 & 1.47 & 0.13 & 0.20 & $4.90\pm0.02$ & 1.43 & 0.38 & 0.00 & $^{1.08}_{0.00}$ & 21.19 \\
H141 & 5.03 & 0.53 & 1.43 & 0.05 & 0.19 & $4.67\pm0.01$ & 2.36 & 0.30 & 0.14 & $^{0.69}_{0.14}$ & 2.21 \\
H143 & 3.75 & 0.62 & 1.75 & 0.11 & 0.22 & $5.11\pm0.02$ & 1.06 & 0.21 & 0.22 & $^{0.43}_{0.22}$ & 10.19 \\
H159 & 5.36 & 0.92 & 1.72 & 0.09 & 0.24 & $4.69\pm0.01$ & 1.47 & 0.32 & 0.29 & $^{0.70}_{0.29}$ & 7.74 \\
H173 & 3.54 & 0.54 & 1.18 & 0.07 & 0.15 & $4.42\pm0.01$ & 1.31 & 0.22 & 0.19 & $^{0.42}_{0.19}$ & 4.91 \\
\hline
Average & 4.38 & 0.87 & 1.71 & 0.41 & 0.22 &  & $1.30$ & $0.22$ & $0.09$ & $0.08$ &  \\
\hline
\multicolumn{12}{p{13.4cm}}{\tiny }\\[-10.0pt]
\multicolumn{12}{p{13.4cm}}{\footnotesize $\dagger$ The parameters $c_2$ and $c_4$ correspond to the slope of the UV extinction, and the far-UV curvature, respectively. For all stars, we adopted $c_1 = 2.030 - 3.007 c_2$; $c_3 = 1.463$; $c_5 = 5.9$, $x_0 = 4.558$, and $\gamma = 0.945$. 
The uncertainties quoted here are statistical uncertainties resulting from the extinction fits; in the case of luminosity, the uncertainties associated with the derivation of stellar parameters (see \citet{2022arXiv220211080B}, their Table A.1.) are not included. \label{tab:app:R136results}}
\end{tabular}
\end{table*}

\newpage
\clearpage 

\section{The extinction law towards R136\label{app:python_equation}}

The \Rmono-dependent spline points that we used for the optical and NIR part of the law can be found in \cref{tab:newsplinesMA14}. A Python function for the computation of the \Rmono-dependent extinction law towards R136 is presented in Listing~\ref{list:python} and can also be found on Github\footnote{\small \url{https://github.com/sarahbrands/ExtinctionR136/}}. 
This function contains both the optical and NIR part as derived by \maiza, but parameterised in the format of \fitz, and the modified UV part (\cref{tab:the_R136_curve}). 

\begin{lstfloat*}
\caption{A Python function that can be used for the computation of the \Rmono-dependent extinction law towards R136. \label{list:python}}
\lstset{linewidth=15cm}
\begin{lstlisting}[language=Python]
import numpy as np
import scipy.interpolate as interpolate

def exctinction_R136(waves,Rv):
    """
    Extinction law tailored to the cluster R136 in the Large Magellanic Cloud,
    where a strong gradient in R5495 values is observed.
    - Shape of the optical and NIR part as in Maiz-Apellaniz et al. (2014),
      but parameterised in a way similar to that of Fitzpatrick (1999).
    - UV part of the curve tailored to R136.
    [Input]
    - waves [numpy array]: wavelength range Angstrom
    - Rv [float]: monochromatic total-to-relative extinction R_5495:
      R5495 = A(lam=5495)/(A(lam=4405)-A(lam=5495))
    [Output]:
    - Alam_Av [numpy array]: normalised extinction as a function of
      wavelength, that is, A(lambda)/A_5495.
      Note that (in broadband equivalents): A(lambda)/A(V) = curve / Rv, with
      curve = k(lambda-V) + Rv = E(lambda-V)/E(B-V) + Rv = Alam / E(B-V)
      This expression is used in e.g. Fitzpatrick et al. (2007).
    """

    # Parameters of the UV part of the extinction curve
    c2 = 1.30                # R136 average
    c1 =  2.030 - 3.007*c2   # As in Fitzpatrick (1999)
    c3 = 1.463               # Bump parameters from Gordon et al. (2003).
    c4 = 0.09                # R136 average
    c5 = 5.9                 # As in Fitzpatrick (1999)
    x0 = 4.558               # Bump parameters from Gordon et al. (2003).
    gamma = 0.945            # Bump parameters from Gordon et al. (2003).

    # Defining inverse wavelength range and the curve components
    xx = 10000./ np.array(waves)
    curve = xx*0.
    xcutuv = 10000.0/2700.0
    xspluv = 10000.0/np.array([2700.0,2600.0])
    iuv = np.where(xx >= xcutuv)[0]
    iopir = np.where(xx < xcutuv)[0]
    if (len(iuv) > 0):
        xuv = np.concatenate((xspluv,xx[iuv]))
    else:
        xuv = xspluv

    # UV part of the curve
    yuv = c1  + c2*xuv
    yuv = yuv + c3*xuv**2/((xuv**2-x0**2)**2 +(xuv*gamma)**2)
    yuv = yuv + c4*(0.5392*(np.maximum(xuv,c5)-c5)**2 +
        0.05644*(np.maximum(xuv,c5)-c5)**3) + Rv
    yspluv  = yuv[0:2]
    if (len(iuv) > 0):
        curve[iuv] = yuv[2::]

    # Optical and NIR spline points
    xsplopir = np.concatenate(([0],10000.0/np.array([26500.0, 18000.0, 12200.0,
        10000.0, 8696.0, 5495.0, 4670.0, 4405.0, 4110.0, 3704.0, 3304.0])))
    ysplopir = np.array((np.polyval([-0.1097 ,0.1195][::-1], Rv),
            np.polyval([-0.2046 ,0.2228][::-1], Rv),
            np.polyval([-0.3826 ,0.4167][::-1], Rv),
            np.polyval([-0.5270 ,0.5740][::-1], Rv),
            np.polyval([-0.6392 ,0.7147][::-1], Rv),
            np.polyval([-0.0002 ,1.0000][::-1], Rv),
            np.polyval([0.7455 ,1.0023][::-1], Rv),
            np.polyval([1.0004 ,1.0000][::-1], Rv),
            np.polyval([1.3149 ,0.9887][::-1], Rv),
            np.polyval([1.7931 ,0.9661][::-1], Rv),
            np.polyval([2.2580 ,0.9689][::-1], Rv)))
    ysplopir = np.concatenate((np.array([0]),ysplopir))

    # If the wavelength range covers optical/NIR, interpolate from UV to optical
    if (len(iopir) > 0):
        tck=interpolate.splrep(np.concatenate((xsplopir,xspluv)),
            np.concatenate((ysplopir,yspluv)),k=3)
        curve[iopir] = interpolate.splev(xx[iopir], tck)

    Alam_Av = curve/Rv

    return Alam_Av
\end{lstlisting}
\end{lstfloat*}

\end{appendix}
\end{document}